\documentclass[aps,twocolumn,superscriptaddress]{revtex4}

\usepackage{graphics,graphicx,dcolumn,bm,fleqn,epic,eepic,float}
\usepackage{amssymb,amsmath,multirow,rotate,color,float}
\usepackage{times}
\definecolor{blue}{rgb}{0,0,1}
\definecolor{red}{rgb}{1,0,0}

\begin{document}
%%%%%%%%%%%%%%%%%%%%%%%%%%%%%%%%%%%%
%\draft
%\preprint{HEP/123-qed}
%%%HEADINGS
\title{Coherence in complex networks of oscillators}

\date{\today}

\author{Pedro G.~Lind}
\homepage[URL: ]{http://www.ica1.uni-stuttgart.de/~lind}
\affiliation{Institute for Computational Physics, 
             Universit\"at Stuttgart, Pfaffenwaldring 27, 
             D-70569 Stuttgart, Germany}
\affiliation{Centro de F\'{\i}sica Te\'orica e Computacional, 
             Av.~Prof.~Gama Pinto 2,
             1649-003 Lisbon, Portugal}
\author{Jason A.C.~Gallas}
\homepage[URL: ]{http://www.if.ufrgs.br/~jgallas}
\affiliation{Institute for Computational Physics, 
             Universit\"at Stuttgart, Pfaffenwaldring 27, 
             D-70569 Stuttgart, Germany}
\affiliation{Instituto de F\'\i sica, Universidade Federal do Rio Grande
             do Sul, 91501-970 Porto Alegre, Brazil}
\author{Hans J.~Herrmann}
\homepage[URL: ]{http://www.ica1.uni-stuttgart.de/~hans}
\affiliation{Institute for Computational Physics, 
             Universit\"at Stuttgart, Pfaffenwaldring 27, 
             D-70569 Stuttgart, Germany}
\affiliation{Departamento de F\'{\i}sica, Universidade Federal do
             Cear\'a, 60451-970 Fortaleza, Brazil}

%%%%%%%%%%%%%%%%%%%%%%%%%%%%%%%%%%%%%%%%%%%%%%%%%%%%%%%%%%%%%%%%%%%%%%
%%%%%%%%%%%%%%%%%%%%% TEXT %%%%%%%%%%%%%%%%%%%%%%%%%%%%%%%%%%%%%%%%%%%
%%%%%%%%%%%%%%%%%%%%%%%%%%%%%%%%%%%%%%%%%%%%%%%%%%%%%%%%%%%%%%%%%%%%%%

%%%%%ABSTRACT
\begin{abstract}
We study fully synchronized (coherent) states in complex networks 
of chaotic oscillators, reviewing the analytical approach of
determining the stability conditions for synchronizability and
comparing them with numerical criteria.
As an example, we present detailed results for
networks of chaotic logistic maps having three different
scale-free topologies:
random scale-free topology, deterministic pseudo-fractal scale-free
network and Apollonian network.
For random scale-free topology we find that the lower boundary 
of the synchronizability region scales approximately as $k^{-\mu}$,
where $k$ is the outgoing connectivity and $\mu$ depends on the local
nonlinearity.  
For deterministic scale-free networks coherence is observed only 
when the coupling is heterogeneous, namely when it is proportional to
some power of the neighbor connectivity.
In all cases, stability conditions are determined from the
eigenvalue spectrum of the Laplacian matrix  and agree well with 
numerical results based on histograms of coherent states in
parameter space. 
Additionally, we show that almost everywhere in the synchronizability
region the basin of attraction of coherent states fills the
entire phase space, and that the transition to coherence is of
first-order.
\end{abstract}

\maketitle

%%%%%%%%
\section{The interplay between dynamics and topology}
\label{sec:intro}

The structure and dynamics underlying complex networks have been
widely investigated, providing insight for many systems where they
arise naturally \cite{livro,albert02,dorogovtsevrev}. 
Complex networks appear in a wide variety of fields ranging
from lasers \cite{meucci02},
granular media \cite{snoeijer04,otto03},
quantum transport \cite{texier04},
colloidal suspensions \cite{tanaka04}, 
electrical circuits \cite{almond04}, and
time series analysis \cite{small02},
to heart rhythms \cite{stewart04},
epidemics \cite{moreno04,dezso02},
protein folding \cite{compiani04}, and
locomotion \cite{zhaoping04} among others
\cite{livro,albert02,dorogovtsevrev}.

From the mathematical point of view, a network is a graph, composed by 
nodes or vertices and by their connections or edges \cite{albert02}.
Sometimes, each node is characterized by some dynamical state
(amplitude), which evolves according to some local contribution and to
the interaction with the neighborhood.
In other words, the complexity of the system underlying the network
may be introduced either in the way nodes are interconnected
(topology) or in the way nodes evolve in time (dynamics).

When studying network dynamics one frequently assumes a regular
topology (lattice) where each node evolves according to some more or
less complicated dynamics, typically fixed points \cite{lind04}, limit 
cycles \cite{strogatz00} or chaotic attractors
\cite{pecora97,anteneodo04}.  
One main goal of this approach is to study the so-called
spatio-temporal chaos which appears in many different spatially
extended systems out from equilibrium, such as hydrodynamical flows,
chemical reactions and biological systems\cite{cross93,kaneko}.
Two main topics in this context concern the study of mechanisms
underlying pattern formation and pattern
selection\cite{cross93,kaneko,wolfram,kaneko93,lind04b} 
and also the study of chaotic synchronization behavior\cite{lind04,sync}. 

Spatially extended systems are fundamentally modeled by
(i) sets of coupled differential equations\cite{cross93} with
nonlinear terms, where both time and amplitude are continuous,  
(ii) cellular automata \cite{wolfram}, where both time and amplitude 
are discrete or
(iii) coupled map lattices\cite{kaneko}, where time is
discrete as in cellular automata, but the space of states is
continuous. 
In all these models the underlying networks have connections whose
range assumes all values from $1$ (nearest neighbors) up to some
maximum range, in particular the size of the system (global coupling
regime). 
In other words, neglecting boundary conditions, these network systems
assume translational symmetry and therefore the underlying network is
called a regular network.

To study more complicated network structures, one usually neglects
node dynamics and all complexity is introduced by the network
topology, i.e.~by the way nodes are connected to each other.
This may be done in three different ways \cite{albert02}: 
by randomly connecting the nodes (random networks \cite{erdos,bollobas}),
by considering some random long-range connections in a regular network
with some small range of couplings (small-world networks
\cite{watts98,strogatz01}), or 
by considering the introduction of new nodes which are connected to the
previous ones following some rule of preferential attachment
(scale-free networks \cite{barabasi99}).
For all these cases there is no translational symmetry and no typical
range connection: connections do not have a `spatial' interpretation.
Therefore, one uses some general topological quantities to
characterize each particular network, namely
the average path length $\langle\ell\rangle$, i.e.~the average minimum
number of connections linking two nodes, 
the clustering coefficient $C$,
defined as the average fraction of neighbors which are connected to
each other, and the distribution of connections $P(k)$,
representing the number of nodes having $k$ connections.
Table \ref{tab1} shows the values of these three quantities for all
three topologies.
%%%%%%%%%%%%%%%%%%%%%%%%%%%%%%%%%%%%%%%%%%%%%%%%%%%%%%%%%%%%%%%%%%%%%%%%
\begin{table}[t]
\centering
\begin{tabular}{cccc}
  \hline\noalign{\smallskip}
%\textbf{Topology}    
                     & \ \ \ \ Random\ \ \ \   
                     & \ \ \ \ \ Small-world\ \ \ \ \  
                     & \ \ \ \ Scale-free\ \ \ \  \\
\noalign{\smallskip}\hline\noalign{\smallskip}
$\mathbf{\langle\ell\rangle}$ &        
          $\ln{N}/\ln{(pN)}$    &
          $N$ for small $p$   &
          $\ln{N}/\ln\ln{N}$   \\
   & &    $\ln{N}$ for large $p$ & \\
\noalign{\smallskip} %\hline\noalign{\smallskip}
$C$  & $\bar{k}/N$   
              & $C_0(1-p)^3$         
              & $\sim N^{-3/4}$   \\
\noalign{\smallskip} %\hline\noalign{\smallskip}
$P(k)$  & $\hbox{e}^{-\bar{k}}\bar{k}^k/k!$       
                 & $\hbox{e}^{-\bar{k}}\bar{k}^k/k!$             
                 & $2m^2/k^3$        \\
\noalign{\smallskip}\hline
\end{tabular}
\caption{Characterizing complex topologies with the topological
  quantities: average path length $\langle\ell\rangle$, clustering
  coefficient $C$ and distributions $P(k)$ of connections $k$.
  Here $N$ is the total number of nodes, $p$ is the probability for
  two nodes to be connected, $\bar{k}$ is the average number of
  connections per node, $C_0$ is the clustering coefficient of the
  regular network from which the small-world network is constructed, and
  $m$ is the number of initial connections of each new node in a
  scale-free network.} 
\label{tab1}
\end{table}
%%%%%%%%%%%%%%%%%%%%%%%%%%%%%%%%%%%%%%%%%%%%%%%%%%%%%%%%%%%%%%%%%%%%%%%%%

Random networks were introduced by Erd\"os and R\'enyi in the late
fifties\cite{erdos} to study organizing principles underlying some
real networks \cite{bollobas}.
In random networks one defines some probability $p(N)$, function of
the total number $N$ of nodes, which determines the probability
for any two nodes to be connected in a total of $N$ nodes. 
Consequently, the connections are typically long-range connections
having a completely irregular structure.
One main goal in studying random networks is to determine the critical
probability $p_c(N)$, beyond which some specific property starts to be
very likely to be observed, e.g.~the critical probability marking a
transition to percolation \cite{christensen98}.  
One important feature of random networks, which also appears in real
networks, is their small average path length $\langle\ell\rangle$,
i.e. the average distance between any two nodes increases slowly with
the system size. However, unlike random networks, real networks also
have large cluster coefficients $C$.

Small-world networks were introduced recently by Watts and Strogatz
\cite{watts98} in order to satisfy both these two features, short
$\langle\ell\rangle$ and large $C$.
Small-world networks have short-range connections between
neighbors, as in regular networks, but they also have long-range
connections similar to random networks, without middle range ones.
There are mainly two procedures to construct a small-world network:
starting from same regular network, where each site is coupled to some
number of first neighbors, one either {\it rewires} each regular
connection with probability $p$ (Watts-Strogatz model \cite{watts98})
or {\it adds} a random connection to each node with probability $p$
(Newman-Watts model\cite{newman99}). 
The second procedure is more appropriate for most purposes, since
it avoids the possibility of generating disconnected
clusters\cite{newman99}.  

Both random and small-world topologies produce typically networks
where connections obey a Poisson distribution (see Tab.~\ref{tab1}). 
However, there are real systems which are scale-free, 
i.e.~where the connection distribution obeys a power-law. 

Scale-free networks were introduced by Barab\'asi and Albert
\cite{barabasi99} using growing and preferential attachment:
one starts with a small amount of nodes totally interconnected, and
adds iteratively one node with $m$ connections to the previous nodes,
chosen from a probability function proportional to their number of
connections. 
With this construction one obtains
analytically\cite{barabasi99b,dorogovtsev00a,kullmann01,krapivsky01}  
a distribution of the connections $P(k)\propto k^{-\gamma}$, where
$\gamma\to 3$ as $N\to \infty$, independently of the initial number of
fully interconnected nodes and of $m$.
%In fact, the value of gamma can be determined analytically
%(i) using continuum theory\cite{barabasi99,barabasi99b} where one
%assumes that the rate of the number of connections of a given node $i$
%equals $mk_i/k_T$ with $k_T$ representing the total number of
%connections in the network at a given iteration, 
%(ii) using a master-equation approach\cite{dorogovtsev00a,kullmann01},
%where one computes the probability for a given node appeared at $t_0$
%to have $k$ connections at time $t>t_0$, or
%(iii) using a rate-equation approach\cite{krapivsky01} where one
%computes the rate of change for the average number $N(k)$ of nodes
%which have $k$ connections. 
It is also possible to generate scale-free networks, by either imposing
{\it a priori} a power-law distribution of all connections randomly
distributed, or by following a deterministic iterative rule for new
nodes. 
The first procedure generates what is usually called a generalized
random graph, while the latter was recently referred as deterministic
scale-free network \cite{barabasi01}.

Deterministic scale-free networks, are hierarchical structures
composed by some succession of generations of nodes, 
i.e.~the set of new nodes appearing simultaneously at a given
iteration during the `construction' of the network,
whose connections follow a particular power-law distribution
\cite{barabasi01,dorogovtsev02,hanspriv,doyecond}, being more easier
to handle.
The main difference between random and deterministic scale-free
networks is due to the local connectivity character of the latter:
random constructions generate irregular long-range connections, while
deterministic networks impose a succession of generation of new nodes
which are, in some way, organized in `space'.
Therefore, deterministic networks are applied for instance in spin
systems \cite{hanspriv}, and geographical and social networks
\cite{hanspriv,gonzalez04}. 

After considering separately dynamical and topological complexity,
the next logical step toward real network dynamics is to consider
them together.
One important question addressed in this context is to know if 
full synchronization between oscillators in such complex topologies 
would appear and under which conditions it is stable.
By full synchronization we mean the convergence of the amplitudes of
all oscillators to the same value, evolving {\it coherently} from then
on. 
Therefore we call henceforth these fully synchronized states {\it
  coherent states}, to distinguish them from {\it partially}
synchronized configurations, when several different clusters of nodes
with the same amplitude are observed \cite{lind04}. 
Synchronization and coherent behavior of oscillator networks with
complex topologies have been studied for the random topology
\cite{chate92,gade96,manrubia99,jost01} and small-world topology
\cite{nishikawa03,barahona02,lago-fernandez00,hong04},
and also scale-free networks \cite{jost01,jostalso,ieee,motter04,lgh}.
In random networks, it is already known \cite{chate92} that with high
coupling strengths it is possible to fully synchronize oscillators 
and the corresponding stability condition may be computed
\cite{gade96} from the matrix of connections characterizing the network.
In small-world networks, synchronizability is observed
\cite{barahona02} only at the end of the small-world regime (high
values of $p$), and recently \cite{nishikawa03} it was found that
heterogeneity in the coupling may destroy coherence. 
These findings for small-world networks are somehow controversial
with the ones of other studies \cite{jost01,lago-fernandez00} and
other topological quantifiers have been proposed \cite{hong04}.
In scale-free networks some recent studies indicate that
synchronizability among oscillators depends on the average connectivity
\cite{motter04} and is robust to delayed flow of information
\cite{jostalso} and to the removal of low-connected nodes \cite{motter04}.

In this manuscript we describe the general approach to study coherent
states in any general complex network of oscillators, and apply it to
the particular case of a scale-free network of discrete-time
oscillators, which is studied in great detail.
We start in Section \ref{sec:stab} by describing the stability analysis
approach to model in Eq.~(\ref{model}) and deduce the corresponding
conditions for synchronizability.
In Section \ref{sec:examp} we apply this stability analysis procedure
to the particular case of scale-free networks, comparing our results
with numerical simulations. 
The random scale-free case is treated in Section \ref{subsec:barabasi},
where we show that the threshold value of such a transition as a
function of coupling strength and outgoing connectivity obeys a
power-law with an exponent that depends on the nonlinearity,
while deterministic scale-free networks are studied in
\ref{subsec:deter}, namely a pseudo-fractal network \cite{dorogovtsev02}
and an Apollonian network \cite{hanspriv,doyecond}. 
Discussion and conclusions are given in Section \ref{sec:discussion}.

%%%%%%%%%%
\section{General approach to analyze coherent states}
\label{sec:stab}

For all the network topologies described above, if one considers
discrete-time oscillators, namely maps of the interval, the equation
of evolution for their amplitudes reads 
\begin{equation}
\vec{x}_{t+1} = \vec{f}(\vec{x}_{t})-\varepsilon\mathbb{G}
                \vec{g}(\vec{x}_{t}) ,
\label{model}
\end{equation}
where $\varepsilon$ is the coupling parameter, $t$ labels time, 
$\vec{x}_t=(x_{t,1},\dots,x_{t,N})$ with $x_{t,i}$ representing the
amplitude at time-step $t$ of node $i=1,\dots,N$, where $N$ is the
total number of nodes, 
$\vec{f}=(f(x_{1}),\dots,f(x_{N}))$ and $\vec{g} = (g(x_{1}),
\dots,g(x_{N}))$ with $f$ and $g$ being real nonlinear functions,
and $\mathbb{G}$ is the coupling (Laplacian) matrix, whose element
$G_{ij}$ represents the relative strength with which node $i$ is
coupled to node $j$, and satisfies the conditions $\sum_{j=1}^N G_{ij}
= 0$ and $G_{ii}=1$ for all $i=1,\dots,N$. In general $\mathbb{G}$ is
a non symmetric matrix. 

Usually, one choses $\vec{g}(\vec{x})\equiv\vec{x}$ when studying
linear coupling, and $\vec{g}(\vec{x})\equiv\vec{f}(\vec{x})$ when
studying nonlinear coupling. Here we consider the nonlinear case.
Apart from this choice, all the information about dynamics is
introduced in function $\vec{f}(\vec{x})$, while all the information
about the coupling topology (regular, random, small-world or
scale-free) and the coupling regime (either homogeneous or
heterogeneous) is included in the coupling matrix $\mathbb{G}$.

From Eq.~(\ref{model}) one easily sees that the coherent state
$x_{t,1}=x_{t,2}=\dots=x_{t,N}=X_t$ evolves in time according to
the local map $X_{t+1}=f(X_t)$. 
There are two ways to study these coherent states: either by studying
the stability of small perturbations of the coherent states or by
making statistics over significant large samples of initial
configurations, counting how many converge to a coherent state.
Some attention to the parameter ranges must be taken, since the basin
of attraction of the coherent states may be bounded by regions of
phase space where amplitudes diverge.
In particular, for maps of the interval one has $0 \le
\varepsilon\le 1$ in order to guarantee convergence of any initial
configuration.  

In this manuscript we will illustrate both analytical and numerical
approaches for the particular case of scale-free networks.
To this end, we define the coupling matrix as $G_{ii}=1$ and
\begin{equation}
G_{ij}=-\frac{k_j^{\alpha}}{\sum_{k\in{\cal K}_i} k_k^{\alpha}}
\label{matC}
\end{equation}
if node $i$ is coupled to node $j$, with $k_j$ representing the number
of nodes of node $j$ and ${\cal K}_i$ is the set of labels of all
neighbors of node $i$. If nodes $i$ and $j$ are not coupled $G_{ij}=0$.
The parameter $\alpha$ is a real number controlling the heterogeneity in
the coupling: positive values of $\alpha$ enhance the coupling
strength with sites having larger number of neighbors, while
negative values favor sites having less neighbors. 
For $\alpha=0$ the coupling between each site and its neighborhood is
homogeneous.

For local dynamics we choose the well-known quadratic map
$f(x)=1-ax^2$, where the free parameter $a$ is restricted to the interval
$-0.25\le a\le 2$ and contains all possible dynamical regimes from a
fixed point (e.g.~$a=0$) to fully developed chaotic orbits (e.g.~$a=2$).

When determining the stability of coherent states, various criteria
are possible. 
For instance, one could compute the maximum Lyapunov exponent and
obtain the conditions where it is negative.
However, maximum Lyapunov exponents do not indicate the existence
of local instabilities in the synchronous state, which may pull the
trajectories apart from the coherent manifold.

The correct approach is based on the variational equation of
Eq.~(\ref{model}) proposed by Pecora and Carroll \cite{pecora98},
which is valid for any network of identical oscillators in what
concerns their local dynamics (quadratic map, Lorenz system, etc.) and
their coupling regime (linear, nonlinear, etc).
For the nonlinear coupling regime, the diagonal form of these
variational equations reads \cite{pecora98,fink00,manrubiabook} 
\begin{equation}
\xi_{t+1,i}=\exp{\left (\Lambda (\varepsilon\lambda_i)\right )}\xi_{t,i}
           =\left [ Df(X)-\varepsilon\lambda_i Df(X)  \right ]\xi_{t,i},
\label{linestab}
\end{equation}
for coherent states $x_{t,i}=X$, where $\Lambda(\varepsilon\lambda_i)$
is the Lyapunov exponent, $Df(X)$ represents the identity matrix
multiplied by the derivative of $f(x)$ computed at $x=X$ and
$\lambda_i$ are eigenvalues of the coupling matrix $\mathbb{G}$.
If $\mathbb{G}$ has zero-sum rows, i.e.~$\sum_{j=1}^N G_{ij}=0 \forall
i$, and all its eigenvalues
$\lambda_1\le\lambda_2\le\dots\le\lambda_N$ are real and nonnegative,
then $\lambda_1=0$ corresponds to the mode parallel to the
synchronization manifold and the largest Lyapunov exponent defines a
master stability function \cite{pecora98}.  
The coherent state is stable whenever $\Lambda(\varepsilon\lambda_i)<0$
for $i=2,\dots,N$ \cite{pecora98,fink00,manrubiabook}.

In our case, it is easy to check from Eq.~(\ref{matC}) that indeed
$\mathbb{G}$ have zero-row sum, yielding $\lambda_1=0$.
Moreover, all the eigenvalues of matrix $\mathbb{G}$ are real
and non-negative, since $\hbox{det}(\mathbb{G}-\lambda\mathbb{I})=
 \hbox{det}(\bar{\mathbb{G}}-\lambda\mathbb{I})$ where
$\bar{\mathbb{G}}$ is a positive semidefinite symmetric matrix, namely 
$\bar{\mathbb{G}}=
      \mathbb{H}^{1/2}\mathbb{K}^{1/2}\mathbb{A}\mathbb{K}^{1/2}
      \mathbb{H}^{1/2}$ with $\mathbb{A}$ being
the adjacency matrix of the network \cite{motter04}, and matrices 
$\mathbb{H}$ and $\mathbb{K}$ being the diagonal matrices with elements
$H_{ii}=1/(\sum_{k\in{\cal K}_i} k_k^{\alpha})$ and
$K_{ii}=k_i^{\alpha}$ respectively. 

From Eq.~(\ref{linestab}) and regarding the ordering of the
eigenvalues $\lambda_i$ one easily concludes that the stability condition
reads
\begin{equation}
\varepsilon_L\equiv\frac{1-\exp{(-\bar{\lambda})}}{\lambda_2}<\varepsilon<
\frac{1+\exp{(-\bar{\lambda})}}{\lambda_N}\equiv\varepsilon_U,
\label{epsint}
\end{equation}
where $\bar{\lambda}$ is the Lyapunov exponent of the local single map.
In particular there is a range of coupling strengths enabling 
synchronizability whenever $\lambda_N/\lambda_2<
(1+\hbox{e}^{-\bar{\lambda}})/(1-\hbox{e}^{-\bar{\lambda}})$ holds.
Therefore, by computing the eigenvalues of the Laplacian matrix
$\mathbb{G}$ one is able to find the range of couplings for which
coherent states are stable.
This approach can be applied for any system ruled by
Eq.~(\ref{model}). 
%%%%%%%%%%%%%%%%%%%%%%%%%%%%%%%%%%%%%%%%%%%%%%%%%%%%%%%%%%%%%%%%%%%%%%%
\begin{figure*}
\begin{center}
\includegraphics*[width=6.7cm]{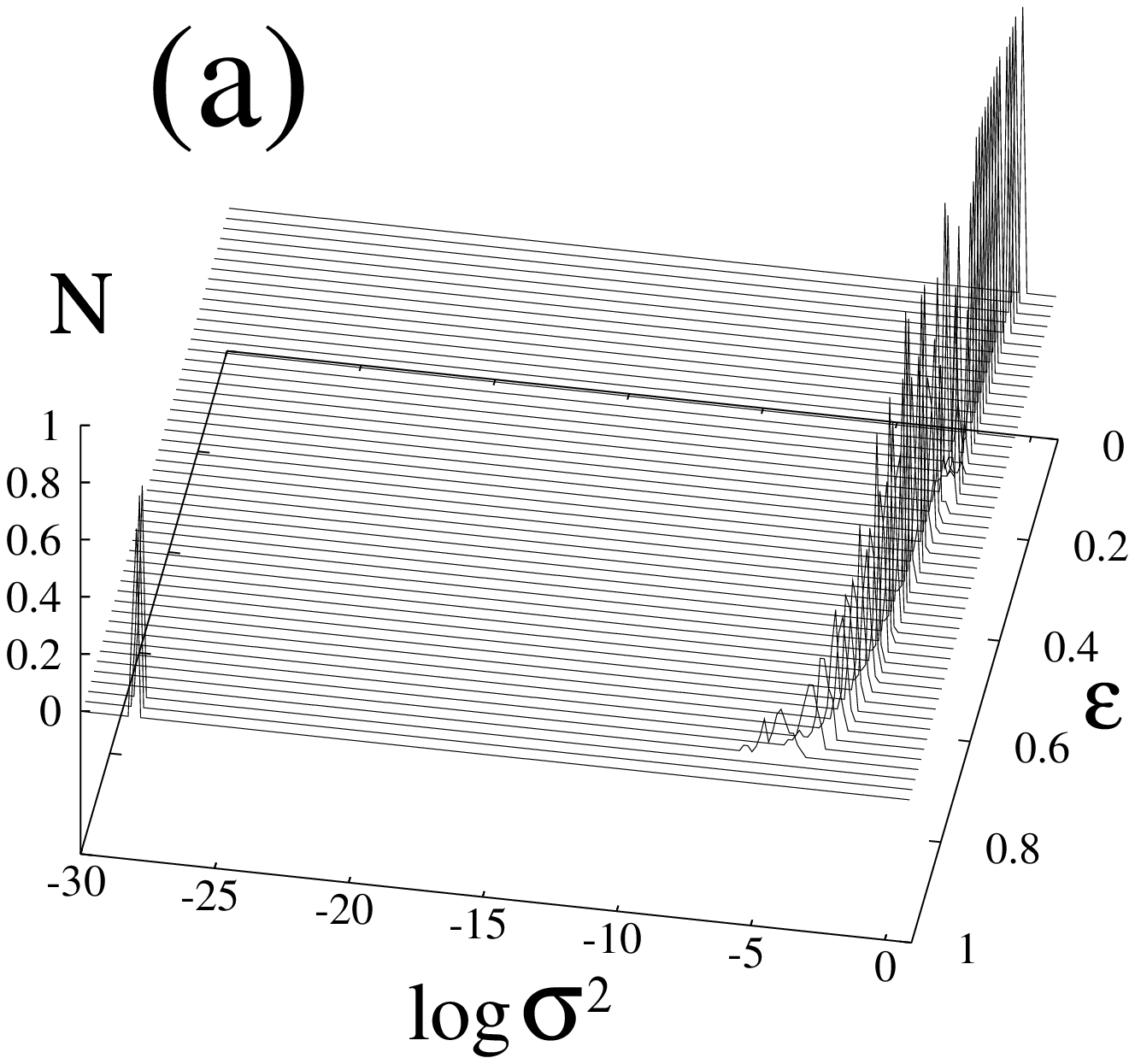}%
\includegraphics*[width=9.0cm]{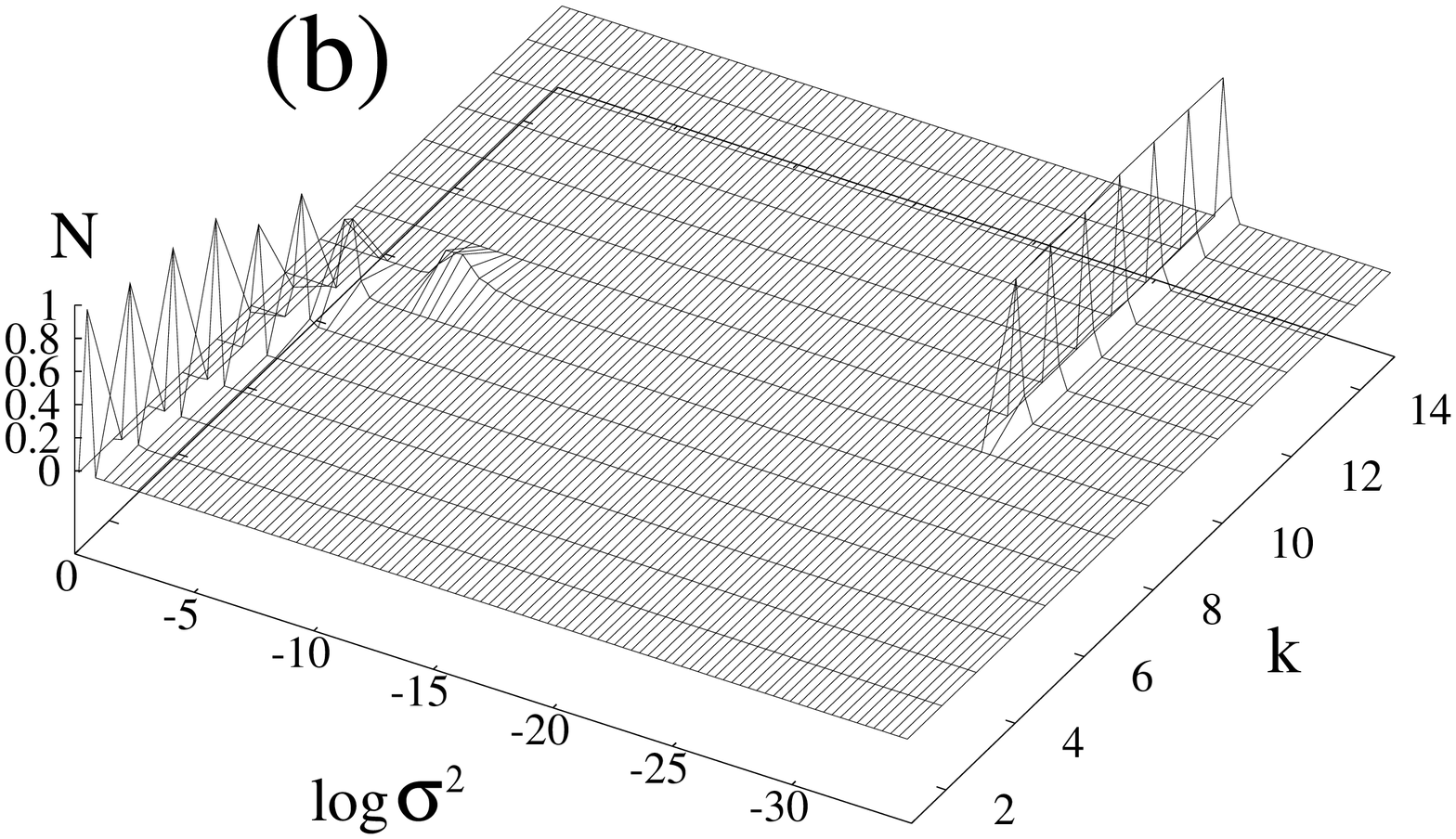}
\end{center}
\caption{\protect 
   Typical histograms of the standard mean square amplitude 
   deviation $\sigma^2$ showing the sharp transition to coherence
   as a function of 
   {\bf (a)} the coupling strength $\varepsilon$ 
             with $k=m_0=8$, and 
   {\bf (b)} the outgoing connectivity $k$ with $\varepsilon=0.95$.
   Values of $N$ represent the fraction of configurations which converge
   to a coherent state from a total of $500$ initial random
   configurations, after discarding transients of $10^4$ time steps. 
   Here $a=2$, $N=1000$, and $\alpha=0$.}
\label{fig1}
\end{figure*}
%%%%%%%%%%%%%%%%%%%%%%%%%%%%%%%%%%%%%%%%%%%%%%%%%%%%%%%%%%%%%%%%%%%%%%%
%%%%%%%%%%%%%%%%%%%%%%%%%%%%%%%%%%%%%%%%%%%%%%%%%%%%%%%%%%%%%%%%%%%%%%%
\begin{figure*}
\begin{center}
\includegraphics*[width=15.5cm]{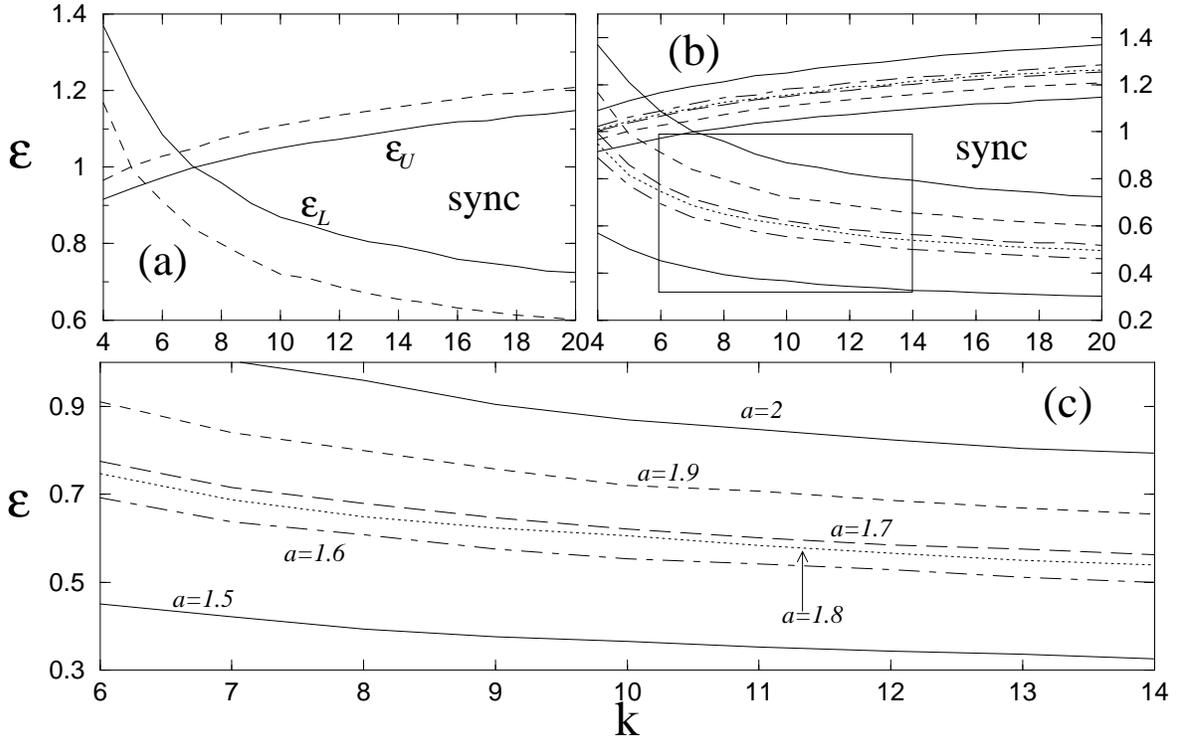}
\end{center}
\caption{\protect 
   Boundary values $\varepsilon_L$ and $\varepsilon_U$ in
   Eq.~(\ref{epsint}) as a function of connectivity $k$ 
   {\bf (a)} for $a=1.9$ (dashed lines) and $a=2$ (solid lines), and
   {\bf (b)} for $a=1.5,1.6,1.7,1.8,1.9$ and $2$, whose inset is
   magnified in {\bf (c)}.
   The regions labeled with 'sync' are the ones where coherent
   solutions are observed, i.e.~$\varepsilon_L < \varepsilon_U$.
   Notice that in (c) the boundary for $a=1.8$ is below the
   one for $a=1.7$, contrary to other values (see text). }
\label{fig2}
\end{figure*}
%%%%%%%%%%%%%%%%%%%%%%%%%%%%%%%%%%%%%%%%%%%%%%%%%%%%%%%%%%%%%%%%%%%%%%%

%%%%%%%%%%
\section{Scale-free networks of coupled logistic maps: an example}
\label{sec:examp}

For the particular case of scale-free networks,
recent results \cite{jost01} show a transition to full
synchronization for two particular values of the nonlinearity $a$ 
in the homogeneous regime ($\alpha=0$), when either the coupling
strength or the number of outgoing connections are varied. 
However, as far as we know there is no detailed study showing how
these coherent states depend on all the parameter models.
Therefore, we present in this Section detailed numerical results
concerning synchronization in oscillator networks with scale-free
topologies. 
Our purpose is to give a complete example of how to study coherent
solutions in complex networks of oscillators, comparing both the
stability analysis and the numerical approaches.

The stability analysis is carried out just by computing the boundary
values $\varepsilon_L$ and $\varepsilon_U$ in Eq.~(\ref{epsint}) as a
function of the model parameters. 
The ranges of values where $\varepsilon_L < \varepsilon_U$
are the ones where coherent solutions appear. As stated above, for
discrete oscillators, ruled by maps of the interval, the condition
$0\le\varepsilon\le 1$ must be added.

Numerically, to detect coherent solutions from a given sample of
initial configurations, we compute the standard mean square deviation
\cite{jost01} 
\begin{equation}
\sigma^2_t=\tfrac{1}{N}\sum_{i=1}^N(x_{t,i}-\bar{x}_t)^2\; , 
\label{sigma}
\end{equation}
where $\bar{x}_t$ is the average amplitude at a 
given time step $t$. 
Whenever $\sigma^2$ is zero within numerical precision,
i.e.~$\sigma^2\sim 10^{-30}$, all the nodes are synchronized at the
same amplitude.

We divide our approach in two parts, the first one concerning
random scale-free networks (Section \ref{subsec:barabasi}) and a
second one concerning deterministic scale-free networks (Section
\ref{subsec:deter}).

%%%%%%%%%%
\subsection{Random scale-free networks}
\label{subsec:barabasi}

In this Section, we use the algorithm of Barab\'asi and
Albert \cite{albert02,barabasi99} to construct the random scale-free
network (see Section \ref{sec:intro}), where at each node one places a
chaotic logistic map.
In a previous work \cite{jost01} a transition to coherence between 
chaotic logistic maps was found for random scale-free networks,
occurring for particularly high coupling strengths, typically of the
order of $\varepsilon_c\sim 0.9$. 
Our simulations have shown that these transitions occur after
discarding transients of $\sim 10^4$ time steps and they do not 
change significantly with the network size.
Moreover, as shown in Fig.~\ref{fig1}, this transition to coherence is
robust with respect to initial configurations, either by varying
the coupling strength $\varepsilon$ (Fig.~\ref{fig1}a) or the outgoing
connectivity $k$ (Fig.~\ref{fig1}b). 
In particular, above the threshold $\varepsilon_c\sim 0.9$, all 
initial configurations converge to a coherent state, indicating
that in this parameter region the basin of attraction of coherent
states fills almost the entire phase space. 

%%%%%%%%%%%%%%%%%%%%%%%%%%%%%%%%%%%%%%%%%%%%%%%%%%%%%%%%%%%%%%%%%%%%%%%
\begin{figure}[b]
\begin{center}
\includegraphics*[width=8.0cm]{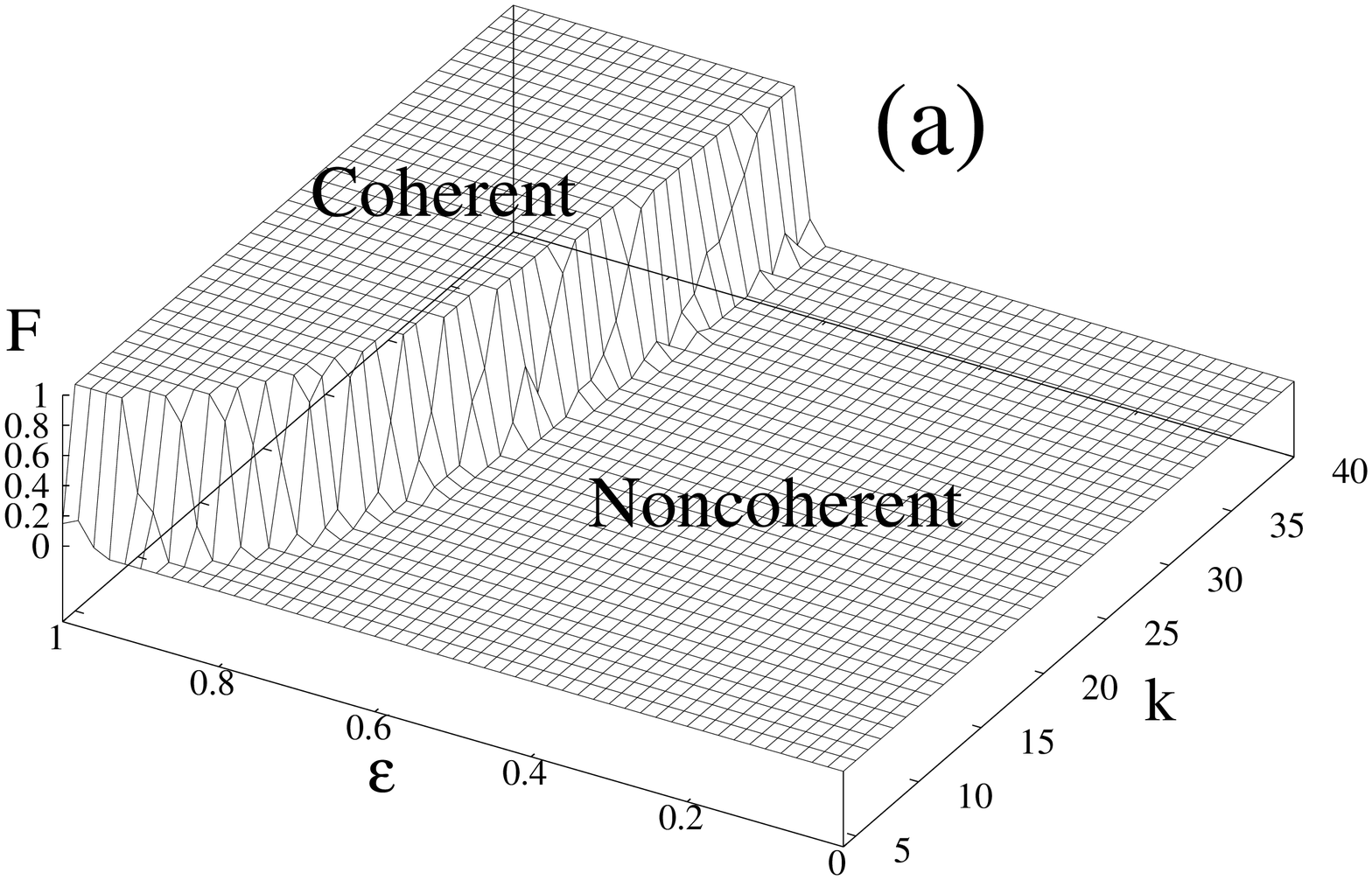}
\includegraphics*[width=8.0cm]{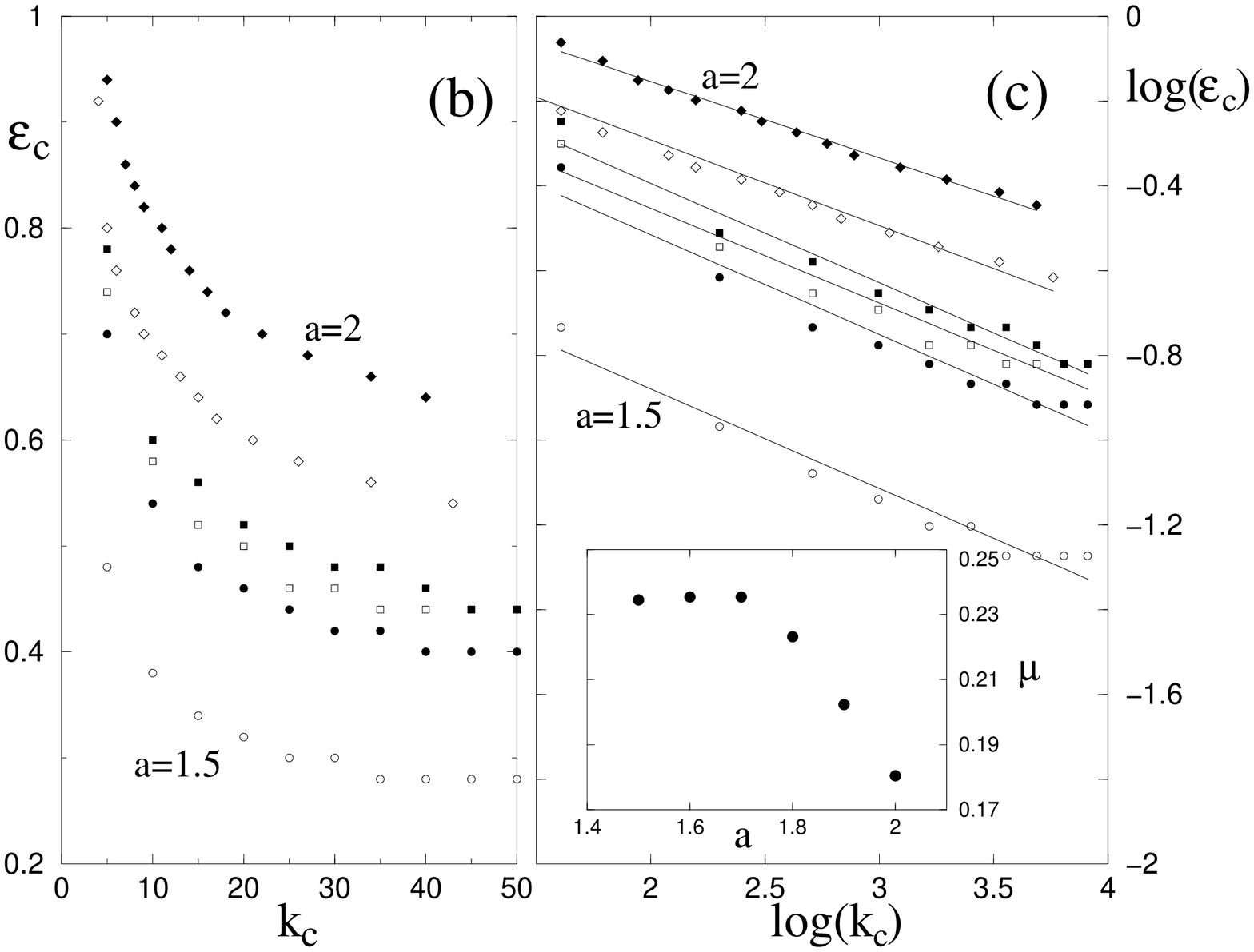}
\includegraphics*[width=8.0cm]{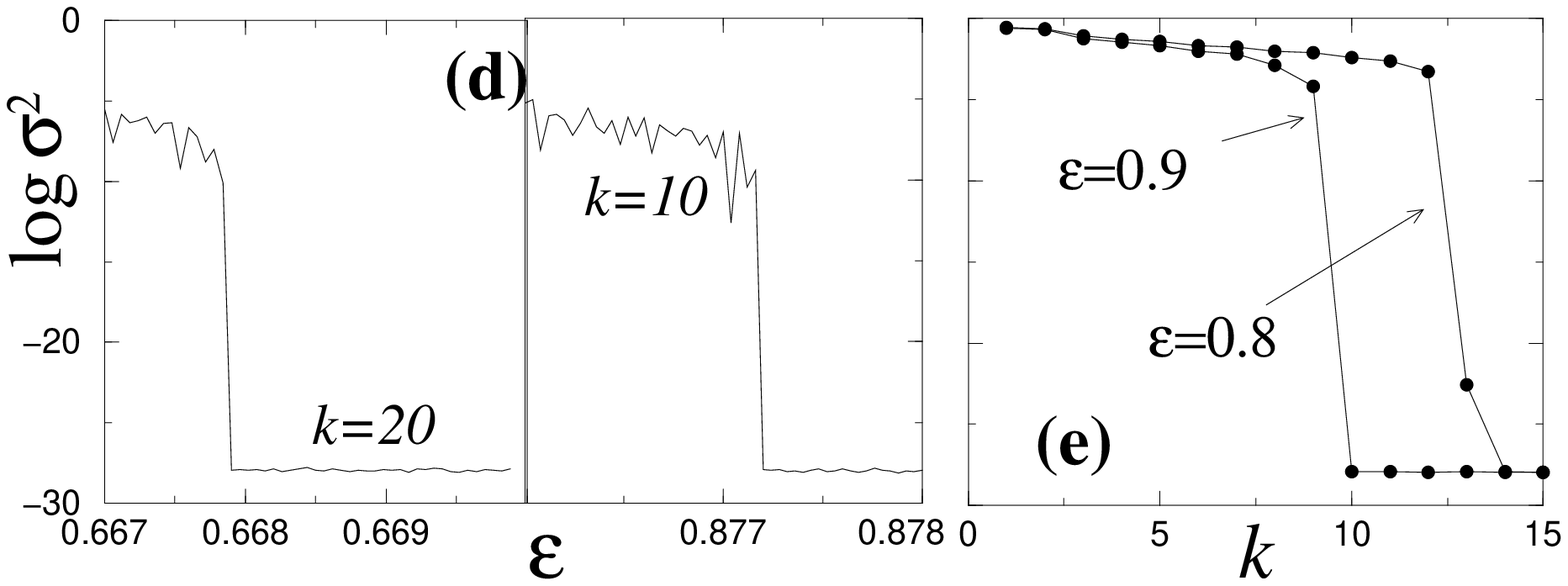}
\end{center}
\caption{\protect 
  Transition to coherence as a function of connectivity $k$ and coupling
  strength $\varepsilon$. 
  {\bf (a)} Fraction $N_{\sigma=0}$ of coherent states from 
  $500$ random initial configurations for $a=2$.
  {\bf (b)} Coherence transition curves in the $(\varepsilon,k)$ plane
  for (from bottom to top) $a=1.5, 1.6, 1.8, 1.7, 1.9$ and $a=2$, and
  {\bf (c)} the same transition in a log-log plot, showing
  power-law dependence between connectivity and coupling strength for
  the transition curves, with an exponent $\mu$ which depends on the
  value of $a$ (see inset).  
  Here $\alpha=0$, $L=1000$ and we used transients of $10^4$ time
  steps.
  By increasing the transient size to $\sim 10^6$ one sees
  clearly that the transition to coherence is of first-order 
  either {\bf (d)} when varying the coupling strength $\varepsilon$ or
  {\bf (e)} when varying the outgoing connectivity $k$.}  
\label{fig3}
\end{figure}
%%%%%%%%%%%%%%%%%%%%%%%%%%%%%%%%%%%%%%%%%%%%%%%%%%%%%%%%%%%%%%%%%%%%%%%
%%%%%%%%%%%%%%%%%%%%%%%%%%%%%%%%%%%%%%%%%%%%%%%%%%%%%%%%%%%%%%%%%%%%%%%
\begin{figure*}
\begin{center}
\includegraphics*[width=15.7cm]{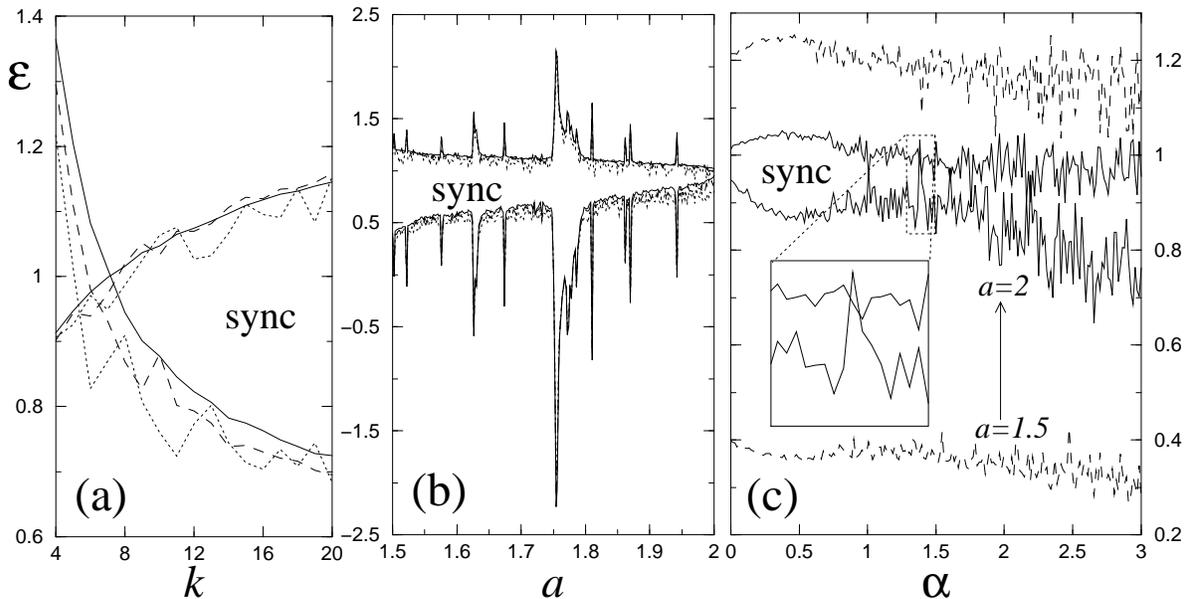}
\end{center}
\caption{\protect 
   Boundary values $\varepsilon_L$ and $\varepsilon_U$ in
   Eq.~(\ref{epsint}) as a function of 
   {\bf (a)} connectivity $k$ with $a=2$ and $\alpha=0$ (solid lines)
   $\alpha=1$ (dashed lines) and $\alpha=2$ (dotted lines),
   {\bf (b)} nonlinearity $a$ for $k=8$ and $\alpha=0,1$ and $2$, and
   {\bf (c)} heterogeneity $\alpha$ with $k=8$ and $a=2$ (solid lines)
   and $a=1.5$ (dashed lines).
   The inset in (c) emphasizes one small region where
   synchronizability is not observed, $\varepsilon_L > \varepsilon_U$
   (see text).
   Here $N=1000$.}
\label{fig4}
\end{figure*}
%%%%%%%%%%%%%%%%%%%%%%%%%%%%%%%%%%%%%%%%%%%%%%%%%%%%%%%%%%%%%%%%%%%%%%%

From stability analysis, we find that in the fully chaotic regime
($a=2$) the transition to coherence occurs for gradually smaller
coupling strength if the connectivity $k$ is increased.
Figure \ref{fig2}a shows the boundaries $\varepsilon_L$ and
$\varepsilon_U$ as a function of $k$ for $a=2$ (solid lines) and
$a=1.9$ (dashed lines) with the same parameter values as in
Fig.~\ref{fig1}. 
As one sees, in both cases the lower boundary $\varepsilon_L$
decreases when $k$ increases, while the upper boundary $\varepsilon_U$
increases beyond $\varepsilon=1$. 
Therefore, one expects that the region of synchronizability increases
for larger values of connectivity $k$.
Figure \ref{fig2}a also shows clearly that for $a=2$ the intersection
between both boundaries, $\varepsilon_L=\varepsilon_U$, occurs just
above $k=7$, which explains why the transition to coherence in
Fig.~\ref{fig1}b occurs at this value.
For $a=1.9$ this transition should occur near $k=5$.
Decreasing even more the nonlinearity coherent solutions are observed
for even smaller connectivities and synchronizability regions increase,
as shown in Fig.~\ref{fig2}b.
To see this feature more clearly we magnify in Fig.~\ref{fig2}c the
inset of Fig.~\ref{fig2}b.
As one sees, one exception occurs for $a=1.8$, where the lower boundary
is {\it below} the one for $a=1.7$, due to the fact that for $a=1.8$
the Lyapunov exponent of the logistic map is smaller than the one for
$a=1.7$, as illustrated below in Fig.~\ref{fig4}.
For all these values of $a$, the single uncoupled map shows chaotic
orbits. %, or at least the orbits have very large periods $\tau > 10^4$.
Moreover, for any other network size $N$, the same curves are obtained.

These analytical predictions extracted from the stability condition in
Eq.~(\ref{epsint}) and shown in Fig.~\ref{fig2} are strongly
corroborated with our numerical simulations as shown in
Fig.~\ref{fig3}. 
In Fig.~\ref{fig3}a we plot the fraction $F$ of initial 
configurations which converge to a coherent state for $a=2$, 
while Fig.~\ref{fig3}b shows the threshold values, 
$\varepsilon_c$ and $k_c$, at the transition
curves where the entire sample of initial configurations converge to a
coherent state, for the same values of $a$ as in Fig.~\ref{fig2}c.
%namely $a=1.5, 1.6, 1.8, 1.7, 1.9$ and $2$.
Here, one clearly sees that there is a clear and sharp transition to
coherence. 
Interestingly, the curves in Fig.~\ref{fig3}b fit very well the ones
in Fig.~\ref{fig2}c, which means that whenever the synchronizability
condition $\varepsilon_L < \varepsilon_U$ holds, coherent states fill
almost entirely the phase space.

Moreover, as illustrated in Fig.~\ref{fig3}c, all curves obey a
power-law, within our numerical precision, 
\begin{equation}
\varepsilon_c\propto k_c^{-\mu} .
\label{powerlaw}
\end{equation}
For the six above values of $a$, the exponents are respectively
$\mu=0.2345$, $0.2354$, $0.2353$, $0.2231$, $0.2023$ and $0.1804$:
the exponent is almost constant below $a\sim 1.7$ and
decreases above this value, as illustrated in the inset of
Fig.~\ref{fig3}c. 

To determine the nature of the transition to coherence seen
in Figs.~\ref{fig3}a and \ref{fig3}b, we plot in Figs.~\ref{fig3}d and
\ref{fig3}e the average standard deviation in the region where
transition to coherence is observed, using much higher resolution.
One clearly sees that the transition to coherence is of first-order,
either when varying $\varepsilon$ or $k$.
That the transitions are indeed of first order is easily recognized by
the clear existence of  hysteresis: 
when increasing either $\varepsilon$ or $k$ the configuration
eventually falls into a coherent state, no longer spontaneously
desynchronizing, no matter how far the parameters are tuned back. 

%As a general first conclusion one could say that, although the 
%exponent $\gamma$ of the power-law distribution of connections
%characterizing scale-free networks does not depend on the outgoing
%connectivity $k$ \cite{barabasi99}, synchronization behavior is quite
%sensitive to this quantity. 
%%%%%%%%%%%%%%%%%%%%%%%%%%%%%%%%%%%%%%%%%%%%%%%%%%%%%%%%%%%%%%%%%%%%%%%
\begin{figure*}
\begin{center}
\includegraphics*[width=8.0cm]{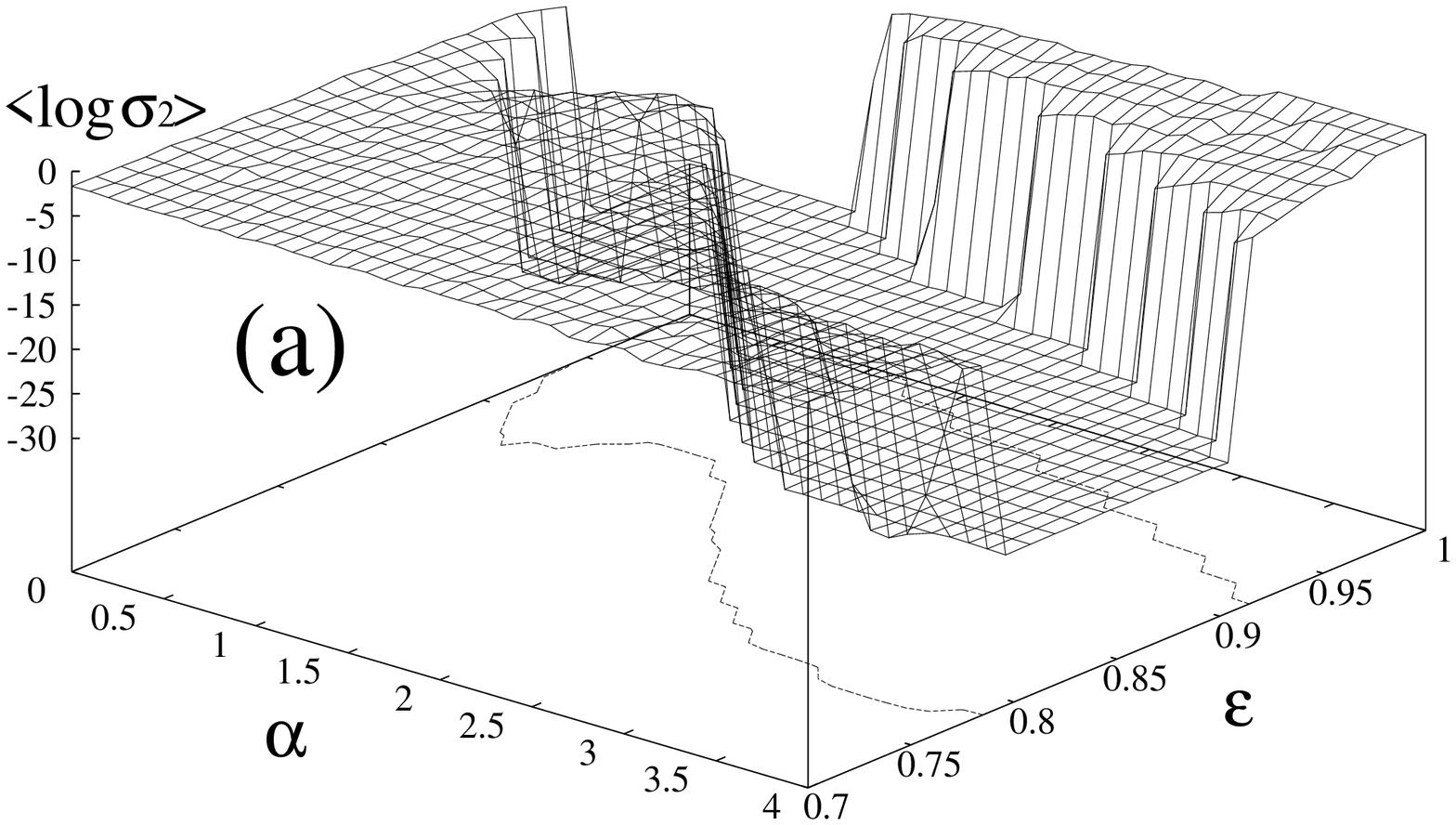}%
\includegraphics*[width=8.0cm]{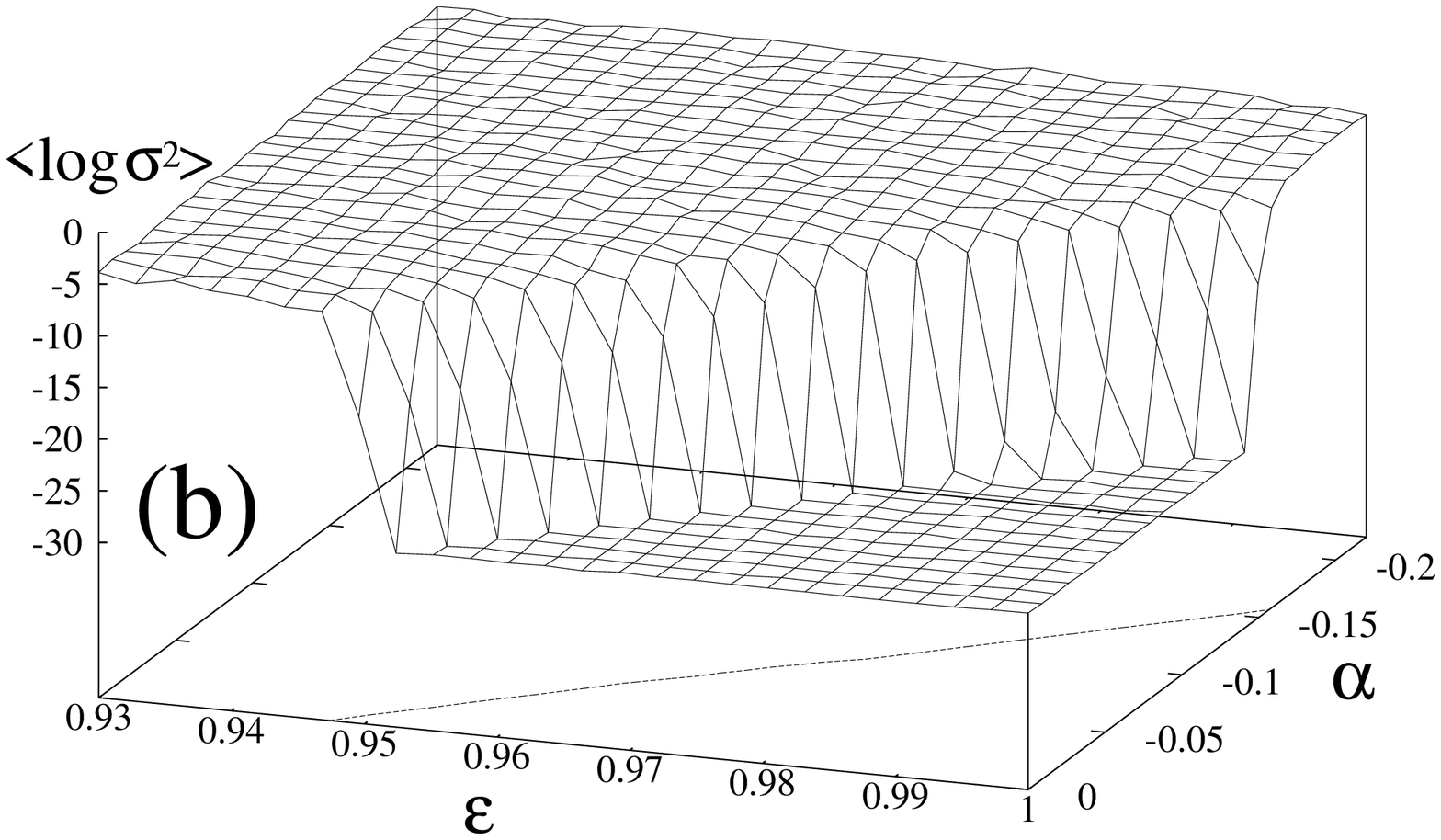}
\end{center}
\caption{\protect
  Transition to coherence as a function of the coupling strength
  $\varepsilon$ and heterogeneity $\alpha$ 
  {\bf (a)} when most connected nodes dominate the dynamics
  ($\alpha>0$) and
  {\bf (b)} when the coupling to nodes with least neighbors is
  strengthened ($\alpha<0$).
  Here, we compute the average standard deviation from a sample of
  $500$ initial configurations and fix $a=2$, $k=m_0=8$ and $N=1000$.}
\label{fig5}
\end{figure*}
%%%%%%%%%%%%%%%%%%%%%%%%%%%%%%%%%%%%%%%%%%%%%%%%%%%%%%%%%%%%%%%%%%%%%%%
%%%%%%%%%%%%%%%%%%%%%%%%%%%%%%%%%%%%%%%%%%%%%%%%%%%%%%%%%%%%%%%%%%%%%%%
\begin{figure*}
\begin{center}
\includegraphics*[width=8.0cm]{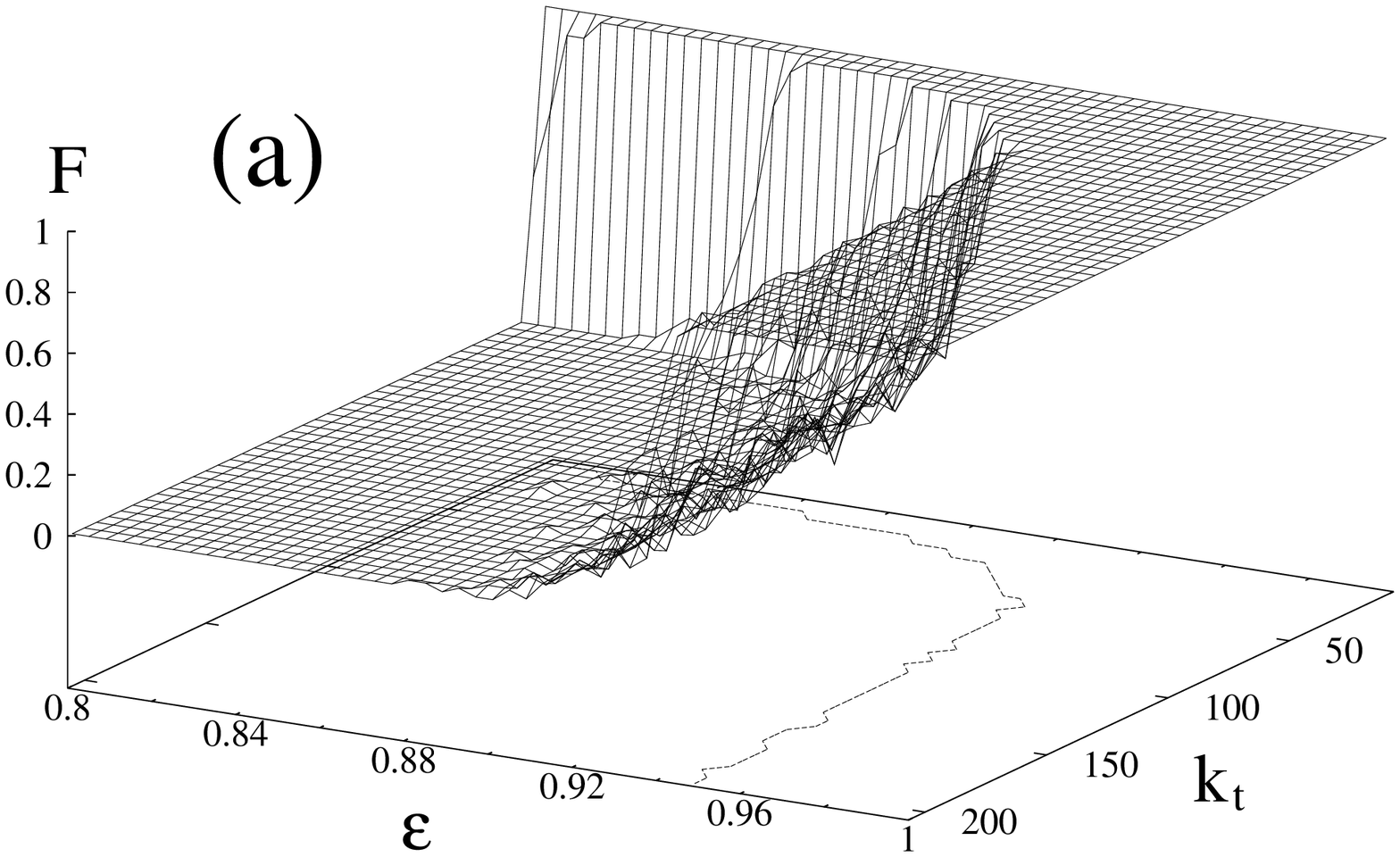}%
\includegraphics*[width=8.0cm]{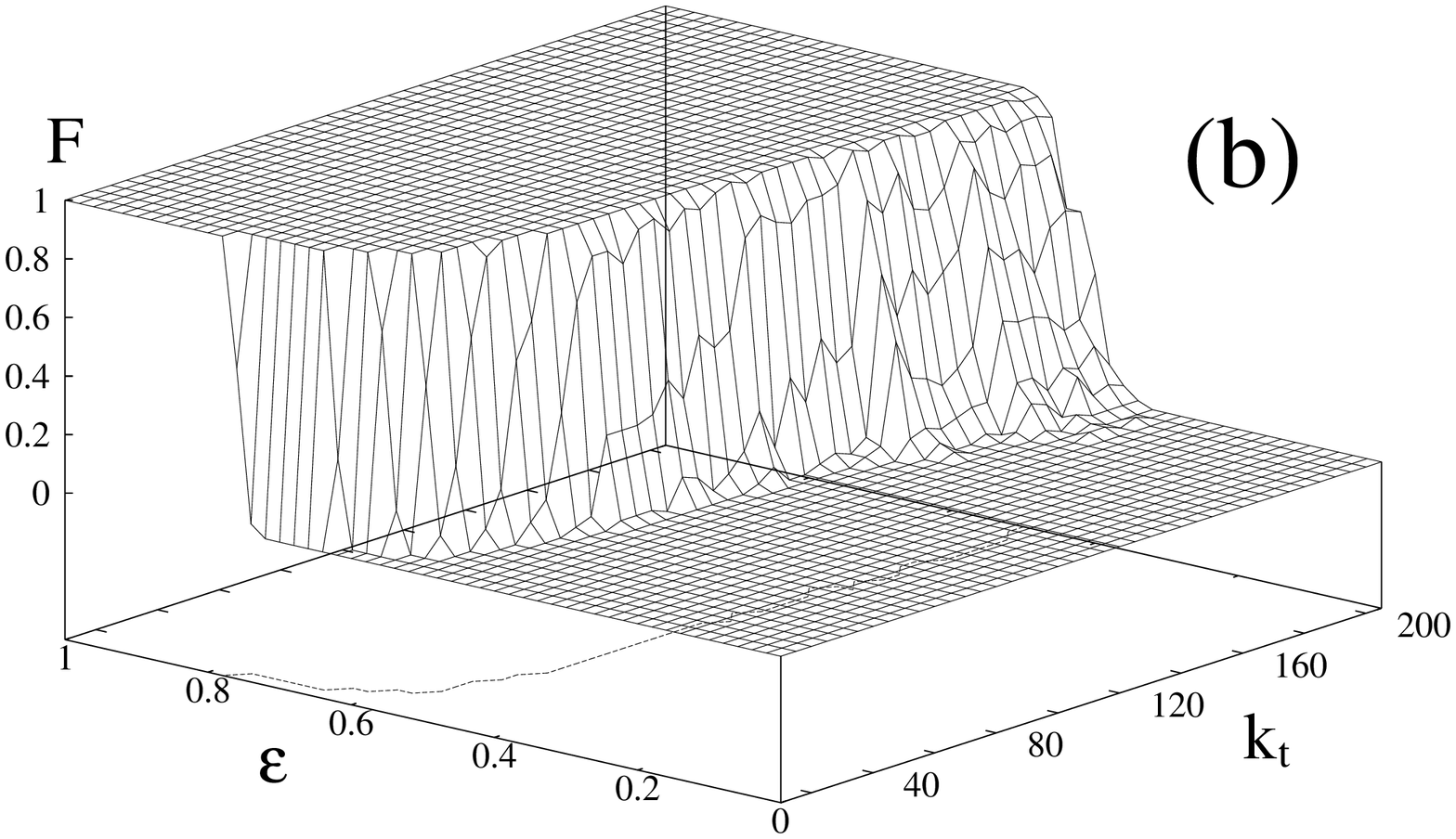}
\end{center}
\caption{\protect 
  Transition to coherence when synchronization is imposed to all nodes
  having a number of neighbors
  {\bf (a)} larger than a threshold $k_t$, and
  {\bf (b)} smaller than $k_t$ (see text).
  Here $a=2$, $k=8$, $\alpha=0$ and $N=1000$.}
\label{fig6}
\end{figure*}
%%%%%%%%%%%%%%%%%%%%%%%%%%%%%%%%%%%%%%%%%%%%%%%%%%%%%%%%%%%%%%%%%%%%%%%

All results till now, concern the case of homogeneous coupling
($\alpha=0$). 
Next, we study the case of heterogeneous coupling.
Figure \ref{fig4} shows the boundaries $\varepsilon_L$ and
$\varepsilon_U$ in Eq.~(\ref{epsint}) as a function of outgoing
connectivity $k$, nonlinearity $a$ and heterogeneity $\alpha$,
covering both the homogeneous and heterogeneous regimes.
Figure \ref{fig4}a shows the two boundaries as a function of $k$
for $a=2$ and $\alpha=0$ (solid lines), $\alpha=1$ (dashed lines) and
$\alpha=2$ (dotted lines).
As one sees for nonzero values of $\alpha$ the boundaries are no
longer smooth curves, but instead they show fluctuations as $k$ is
increased, enlarging and shrinking alternately the region of
synchronizability, labeled as 'sync'.
When varying $a$ (Fig.~\ref{fig4}b) the boundaries
are mainly controlled by the Lyapunov exponent of the
local map (see Eq.~(\ref{epsint})), where $\varepsilon_L$
(resp.~$\varepsilon_U$) decreases (resp.~increases) whenever a
periodic window occurs \cite{ottbook}.
The fluctuations observed in Fig.~\ref{fig4}a are clearly seen in
Fig.~\ref{fig4}c, where the stability boundaries are plotted as a
function of $\alpha$ fixing $k=8$ and $a=2$ (solid lines) and $a=1.5$
(dashed lines). 
The fluctuations are much higher for $\alpha > 1$ and for the fully
chaotic regime both boundaries may even cross each other suppressing
synchronizability (see inset of Fig.~\ref{fig4}c).
Moreover, the lower boundary $\varepsilon_L$ decreases till
$\alpha\sim 0.5$, then increases till $\alpha\sim 1$ and decreases in
average from there on.

All these analytical results computed from Eq.~(\ref{epsint}) and
matrix $\mathbb{G}$ in Eq.~(\ref{model}) are corroborated by our
numerical simulations.
In particular, the boundaries $\varepsilon_L$ and $\varepsilon_U$ seen
in Fig.~\ref{fig4}c are obtained also when plotting the contour of
Fig.~\ref{fig5}a, where we plot the average standard 
deviation from a sample of $500$ initial configurations and vary the
coupling strength and heterogeneity for $a=2$ and $L=1000$.
While Fig.~\ref{fig5}a shows the numerical results for $\alpha > 0$,
i.e. in the case where nodes are more strongly coupled to the
neighbors with higher connectivities, Fig.~\ref{fig5}b shows the 
transition to coherence when $\alpha < 0$.
Here synchronizability is observed only for $\alpha \gtrsim -0.15$ and
for very high coupling strengths $\varepsilon\gtrsim 0.95$.
%Moreover, this transition curve is obtained also from the eigenvalues
%of matrix $\mathbb{G}$ which varies linearly with heterogeneity
%$\alpha$. 

We end our study of coherent solutions in random scale-free networks
by investigating briefly the role of hubs in the lattice.
Instead of strengthening the coupling to the most connected nodes by 
increasing $\alpha > 0$, we now fix $\alpha=0$ and impose
synchronization between all the nodes with more than a certain
threshold $k_t$ of neighbors and observe which fraction of the initial
configurations converges to a coherent state.
In this case the transition to coherence converges asymptotically to a
limit of the coupling strength, as shown in Fig.~\ref{fig6}a.
The same occurs when synchronization is imposed to all nodes with {\it
  less} than $k_t$ neighbors, as shown in Fig.~\ref{fig6}b.

%%%%%%%%%%
\subsection{Deterministic scale-free networks}
\label{subsec:deter}

\begin{figure}[htb]
\begin{center}
\includegraphics*[width=4.2cm]{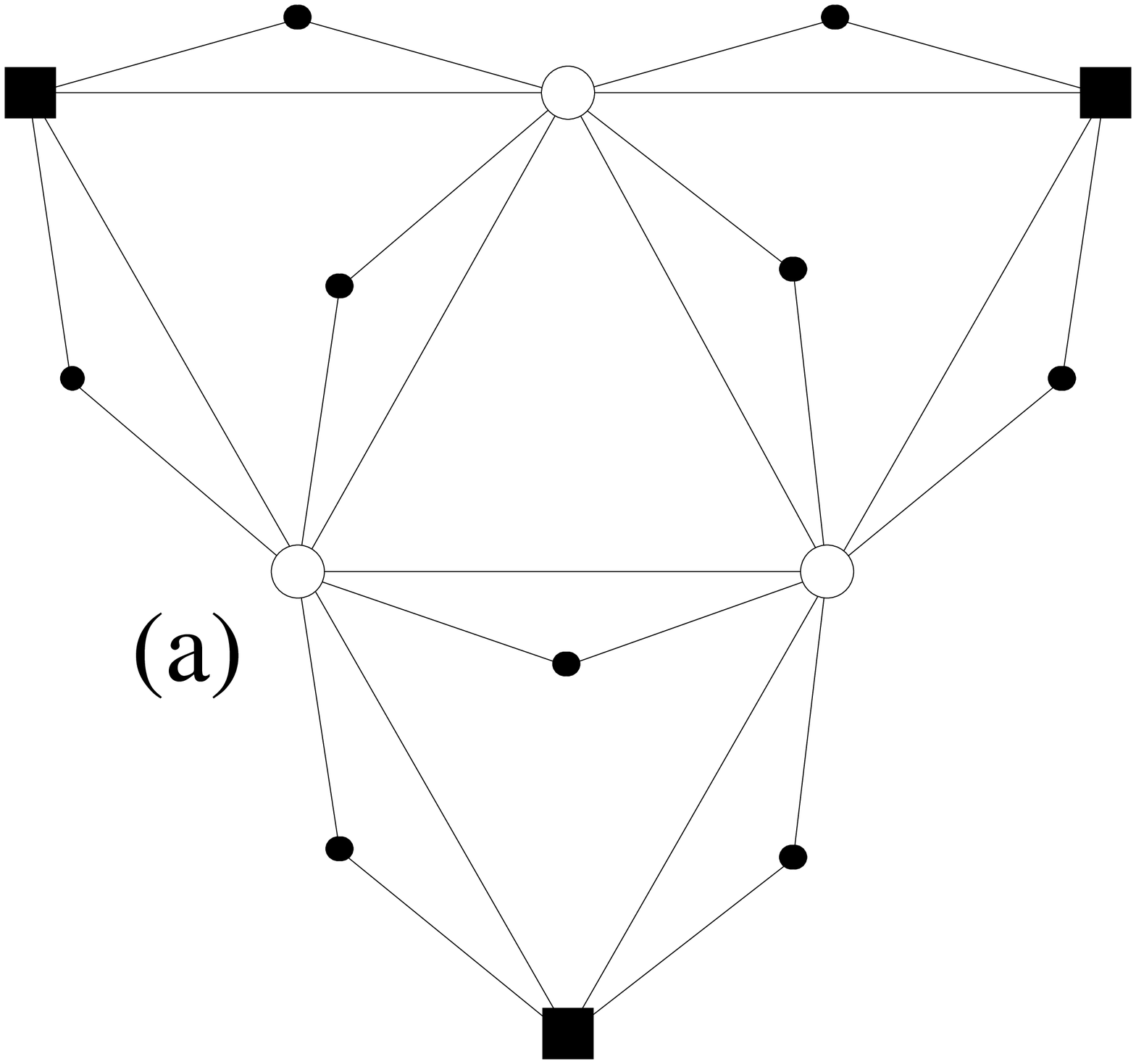}%
\includegraphics*[width=4.2cm]{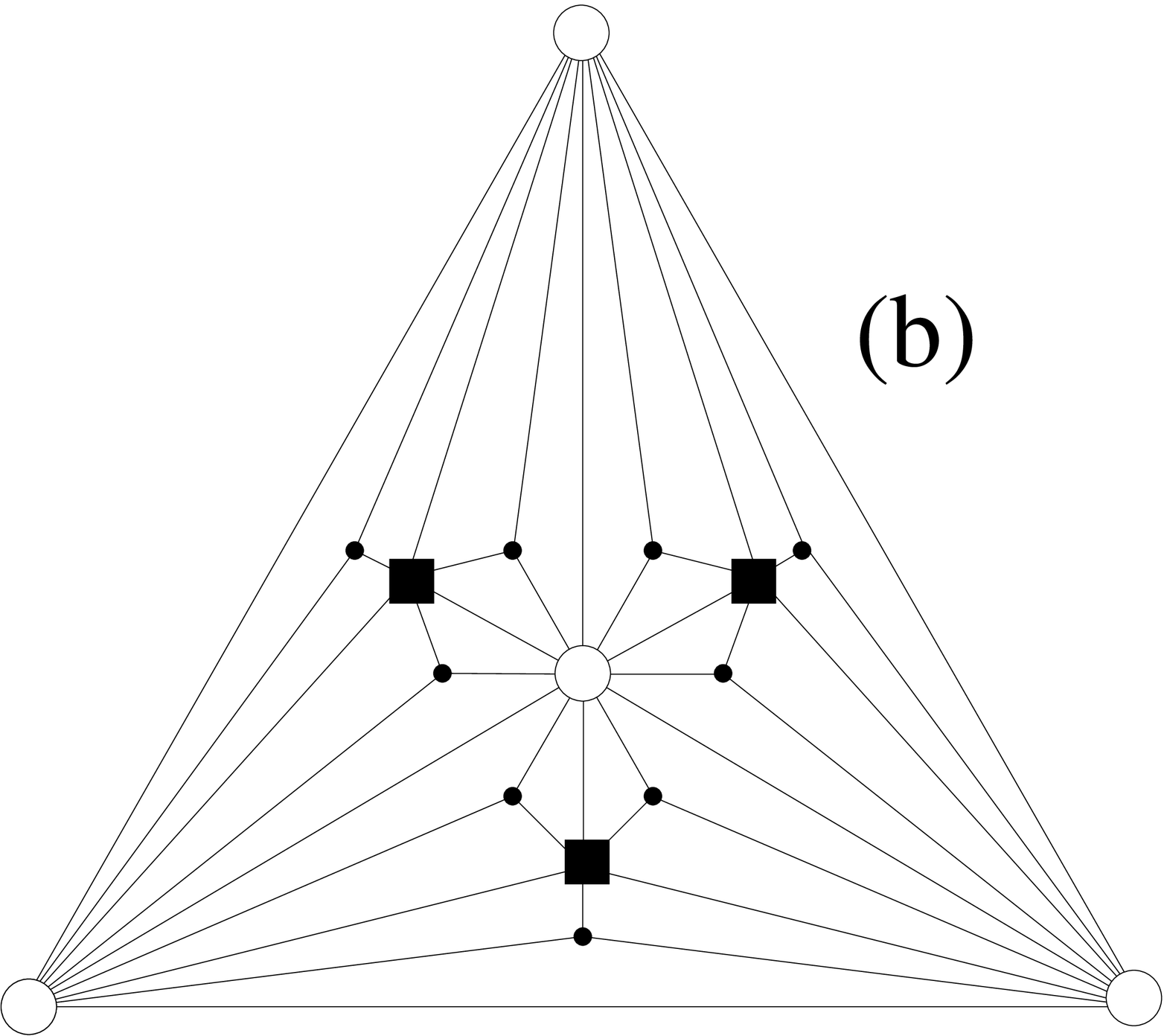}
\end{center}
\caption{\protect 
   Illustrations of two deterministic scale-free networks:
   {\bf (a)} the pseudo-fractal network \cite{gonzalez04}, and
   {\bf (b)} the Apollonian network \cite{hanspriv}.
   Identical symbols label nodes belonging to the same generation $n$
   (see text), namely $\bigcirc$ for $n=0$, $\blacksquare$ for $n=1$
   and $\bullet$ for $n=2$.} 
\label{fig7}
\end{figure}

In the previous Section we focused on random scale-free networks,
i.e.~growing networks where new nodes are connected following
probabilistic rules.  
In this Section we study deterministic scale-free networks
\cite{barabasi01,dorogovtsev02,hanspriv},
using two different deterministic topologies:
the pseudo-fractal scale-free network introduced by Dorogovtsev et al
\cite{dorogovtsev02} and the Apollonian network introduced by Andrade et
al \cite{hanspriv} and studied also in Ref.~\cite{doyecond}.
Both networks are illustrated in Fig.~\ref{fig7}.
%%%%%%%%%%%%%%%%%%%%%%%%%%%%%%%%%%%%%%%%%%%%%%%%%%%%%%%%%%%%%%%%%%%%%%%
%%%%%%%%%%%%%%%%%%%%%%%%%%%%%%%%%%%%%%%%%%%%%%%%%%%%%%%%%%%%%%%%%%%%%%%

The pseudo-fractal network of Dorogovtsev is obtained, starting from
three interconnected nodes, and at each iteration
each edge generates a new node, attached to its two vertices.
Figure \ref{fig7}a illustrates this network after three iterations, i.e.~with
three generations of nodes. 
The number of nodes $N_n$ and the number of connections $V_n$
increases with the number of generations as
\cite{dorogovtsev02}
\begin{subequations}
\begin{eqnarray}
N_n &=& \tfrac{3}{2}(3^n+1)\; ,\label{pseudoL}\\
V_n &=& 3^{n+1}\; .\label{pseudoM}
\end{eqnarray}
\end{subequations}
From Fig.~\ref{fig7}a one easily sees that this network has indeed a
scale-free topology, since the number of nodes with degree
$k=2,2^2,\dots,2^{n-1},2^n$ and $2^{n+1}$ is equal to
$3^n,3^{n-1},\dots,3^2,3$ and $3$ respectively.
In particular, the exponent of this power-law distribution is
$\gamma=1+\ln{3}/\ln{2}$.
Moreover, the cluster coefficient of a node with degree $k$ is
$C=2/k$, and the average path length is approximately
$\langle\ell\rangle \simeq 4\ln{N_n}/(9\ln{3})$.

The Apollonian network is constructed in a different way:
one starts with three interconnected nodes, defining a triangle;
at $n=0$ one puts a new node at the center of the triangle and joins
it to the three other nodes, thus defining three new smaller triangles;
at iteration $n=1$ one adds at the center of each of these three triangles
a new node, connected to the three vertices of the triangle, defining
nine new triangles and so on (see Fig.~\ref{fig7}b). 
The number of nodes and the number of connections are given respectively by
\begin{subequations}
\begin{eqnarray}
N_n &=& \tfrac{1}{2}(3^{n+1}+5)\; ,\label{apollonL}\\
V_n &=& \tfrac{3}{2}(3^{n+1}+1)\; .\label{apollonM}
\end{eqnarray}
\end{subequations}
The distribution of connections obeys a power-law, since the number of
nodes with degree $k=3,3\cdot 2,3\cdot 
2^2,\dots, 3\cdot 2^{n-1},3\cdot 2^{n}$ and $2^{n+1}$ is equal to
$3^n,3^{n-1},3^{n-2},\dots,3^2,3,1$ and $3$ respectively, and the
exponent $\gamma$ is the same as for the pseudo-fractal network.
Moreover, a node with $k$ neighbors has a cluster coefficient of
$C\simeq 4/k$ as reported in \cite{doyecond}, converging on average
to $C_{\infty}=0.828$, and the average path length grows weaker than
$\ln{N_n}$\cite{hanspriv}. 

%From Fig.~\ref{fig7} and the description above, one easily concludes
%that for the pseudo-fractal network the outgoing connectivity is fixed
%at $k=2$, while for Apollonian networks one has $k=3$.
Despite that both networks have similar values for the topological
quantities, they are quite different from the geometrical point of view:
the pseudo-fractal network has no metric, while the Apollonian
network is embedded in Euclidean space and fills it densely as
$n\to\infty$, being particularly suited for describing geographical
situations \cite{hanspriv}. 

For stability analysis purposes (see Section \ref{sec:stab}), 
the Laplacian matrix $\mathbb{G}$ of deterministic networks can be
analytically determined from the adjacency matrix $\mathbb{A}=\{
a_{ij}\}$, since they are related by
\begin{equation}
\mathbb{G} = \mathbb{I}+\mathbb{A}\mathbb{T} ,
\end{equation}
where $\mathbb{I}$ is the identity matrix and the values of matrix
$\mathbb{T}=\{T_{ij}\}$ are defined by
\begin{equation}
T_{ij} = -\frac{a_{ji}\left [ \sum_{k=1}^N a_{ik}\right ]^{\alpha}}
              {\sum_{p=1}^N a_{jp}\left [ \sum_{k=1}^N a_{pk}\right
              ]^{\alpha}}.
\end{equation}
%As mentioned by Barab\'asi et al \cite{barabasi01}, a strong
%advantage of deterministic networks is that it is often possible to
%compute {\it analytically} their properties, for example the adjacency
%matrix, whose eigenvalue spectrum characterizes the topology
%\cite{albert02}.
A simple way to write the adjacency matrix of the pseudo-fractal
network is
\begin{equation}
\mathbb{A}_{n} =
\left [
\begin{array}{cc}
\mathbb{A}_{n-1}  & \mathbb{M}_{n-1} \\
\mathbb{M}_{n-1}^T  & \emptyset
\end{array}
\right ]_{N_n \times N_n}  \; ,
\label{pseudo_adjmat}
\end{equation}
where $N_n$ is given by Eq.~(\ref{pseudoL}), $\mathbb{M}^T$ represents
the transposed matrix of $\mathbb{M}$ and for each generation
$n=1,2,\dots$ the matrix $\mathbb{M}_n$ reads
\begin{equation}
\mathbb{M}_n =
\left [
\begin{array}{ccc}
\mathbb{M}_{n-1} & \mathbb{M}_{n-1} & \emptyset \\
\emptyset      & \emptyset      & \mathbb{B}_{n-1}
\end{array}
\right ]_{2\cdot3^{n-1}\times 3^n}  \; ,
\end{equation}
with
\begin{equation}
\mathbb{B}_{n-1} =
\left [
\begin{array}{cccc}
\mathbb{A}_0 & \emptyset  &  \dots  & \emptyset \\
\emptyset  & \mathbb{A}_0 &  \dots  & \emptyset \\
\vdots     & \vdots      &  \ddots & \vdots    \\
\emptyset  & \emptyset  & \dots   & \mathbb{A}_0
\end{array}
\right ]_{3^{n-1}\times 3^{n-1}}   
\end{equation}
and whose starting form is
\begin{equation}
\mathbb{M}_0 = \mathbb{A}_0 =
\left [
\begin{array}{ccc}
0 & 1 & 1 \\
1 & 0 & 1 \\
1 & 1 & 0
\end{array}
\right ]_{3\times 3}  \; .
\end{equation}

For the Apollonian network, the adjacency matrix is given by the same 
recurrence of Eq.~(\ref{pseudo_adjmat}), but this time with
\begin{equation}
\mathbb{A}_0 =
\left [
\begin{array}{cccc}
0 & 1 & 1 & 1\\ 
1 & 0 & 1 & 1\\ 
1 & 1 & 0 & 1\\ 
1 & 1 & 1 & 0
\end{array}
\right ]\; ,
\label{a0_apollon_adjmat}
\end{equation}
and $\mathbb{M}_n$ being a matrix with $(3^n+5)/2$ rows and $3^n$ columns
and having in each column only three non-zero elements.

\begin{figure*}
\begin{center}
\includegraphics*[width=15.0cm]{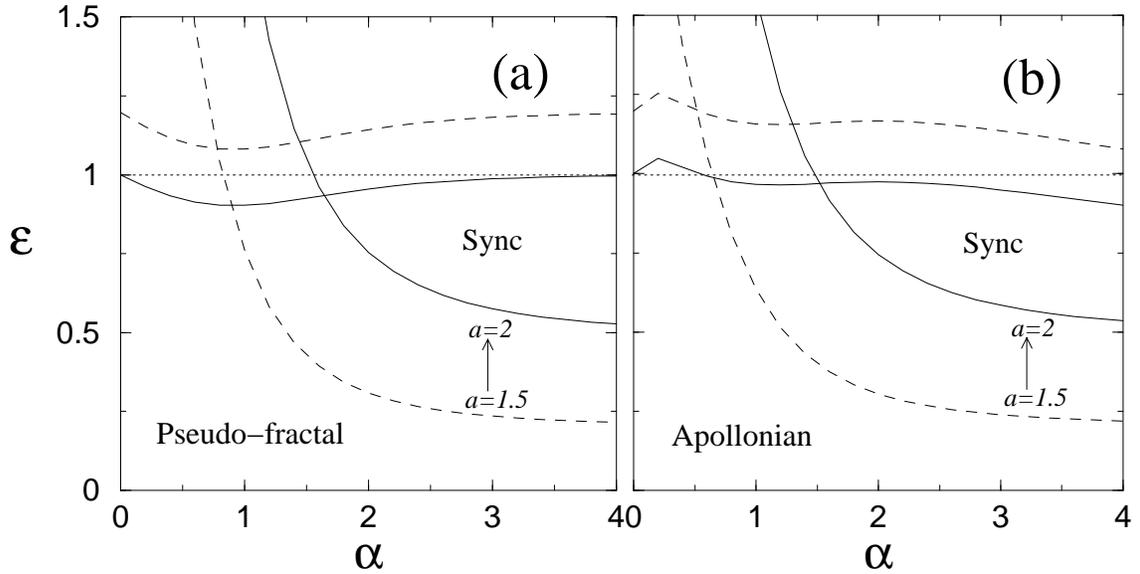}
\end{center}
\caption{\protect 
   Boundary $\varepsilon_L$ and $\varepsilon_U$ for synchronizability
   as a function of heterogeneity $\alpha$ for
   {\bf (a)} the pseudo-fractal network and
   {\bf (b)} the Apollonian network, with
   $a=2$ (solid lines) and $a=1.5$ (dashed lines).
   For each network we use $6$ generations of nodes (see text).
   These eigenspectra are the same for any number of generations.}
\label{fig8}
\end{figure*}

\begin{figure*}
\begin{center}
\includegraphics*[width=7.5cm]{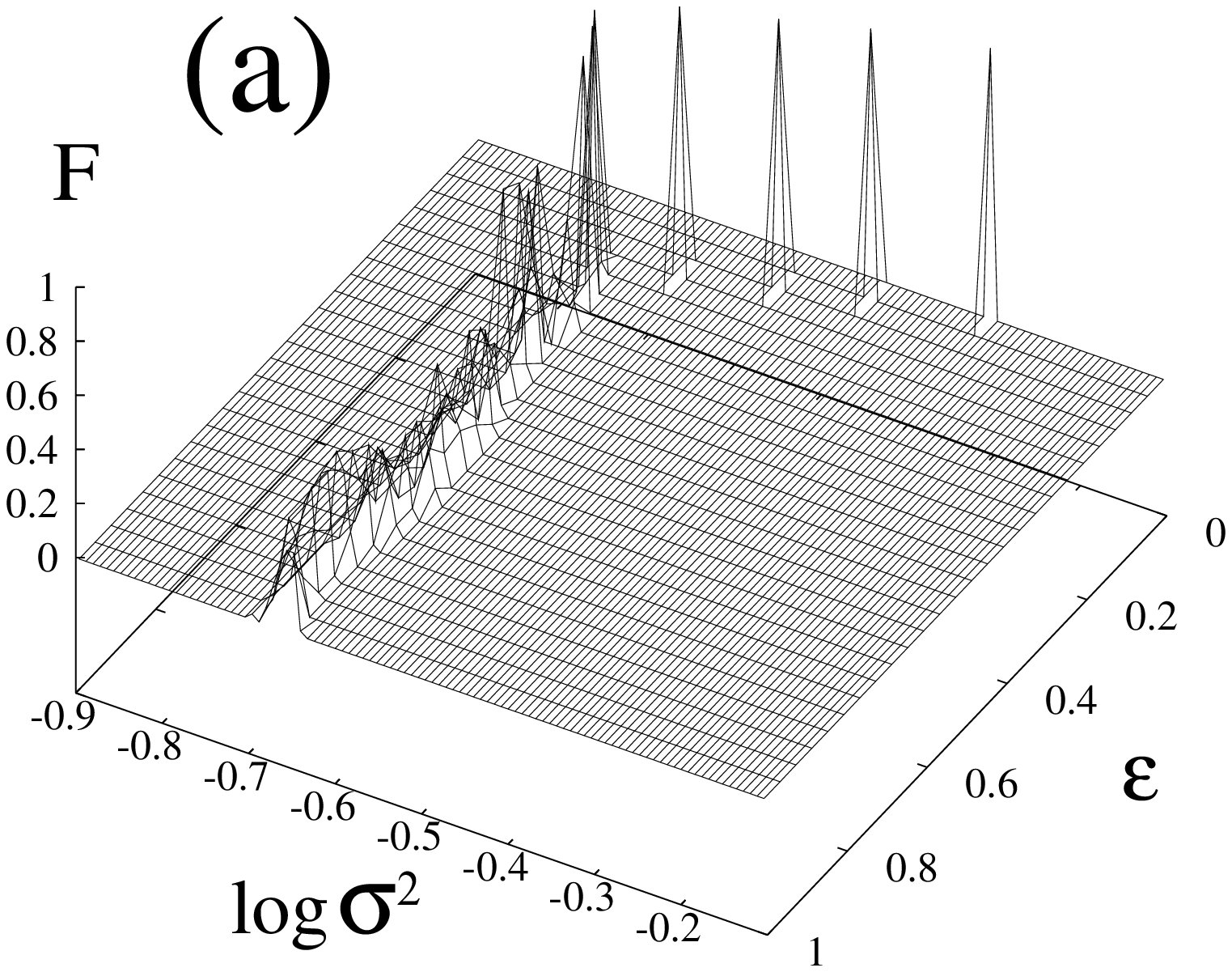}%
\includegraphics*[width=7.5cm]{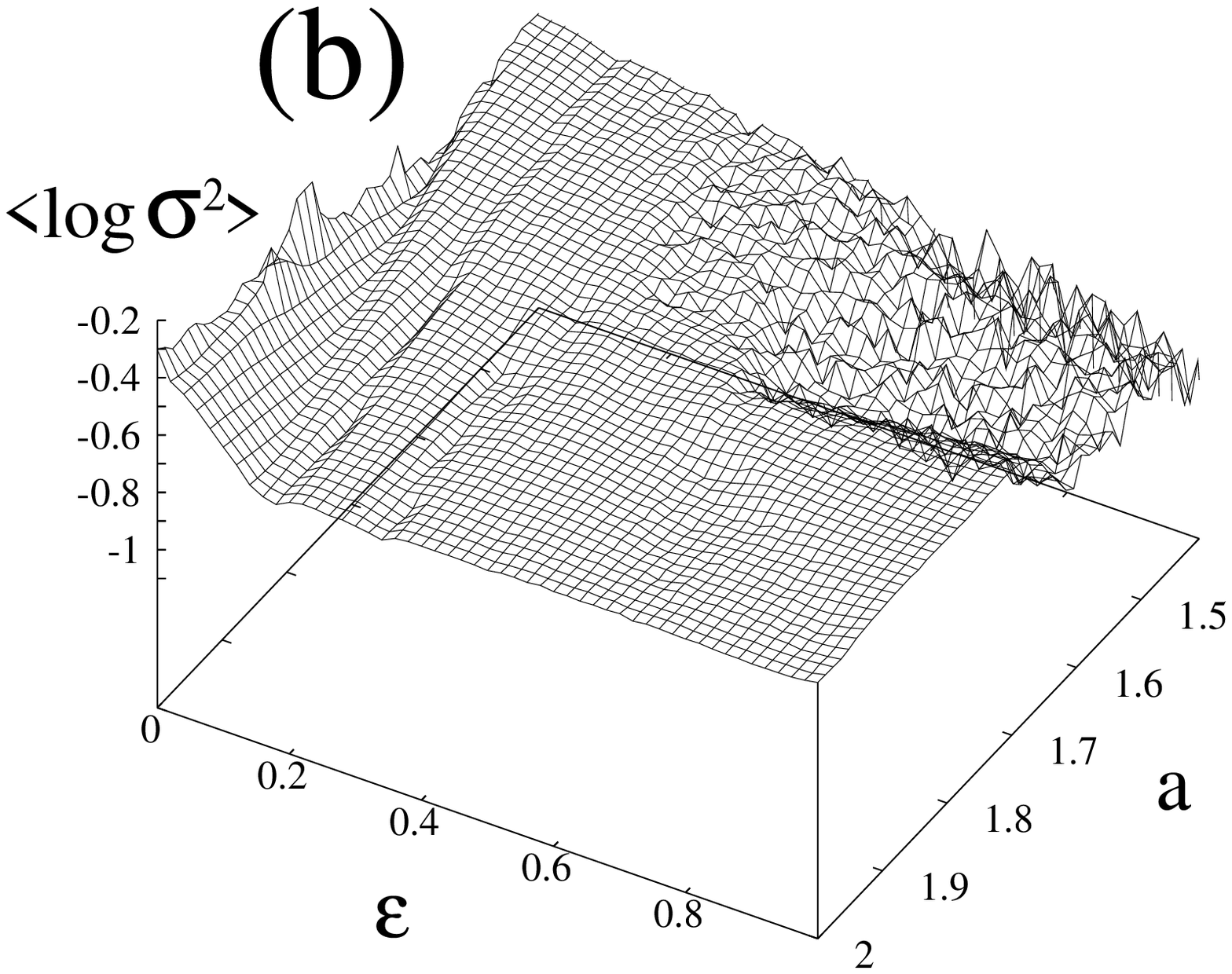}
\end{center}
\caption{\protect 
   {\bf (a)}
   Typical histogram of the standard mean square deviation $\sigma^2$ for
   the pseudo-fractal network as a function of the coupling strength
   $\varepsilon$, with $a=2$ and $\alpha=0$.  
   A similar result is obtained for the Apollonian network.
   {\bf (b)} Histogram of the standard mean square deviation
   $\sigma^2$ as a function of nonlinearity $a$ and coupling strength
   $\varepsilon$, for deterministic scale-free networks with $\alpha=0$.
   The mean square deviation is averaged over a sample of $500$ initial
   configurations and during $100$ time steps, after discarding
   transients of $10^4$ time steps.}
\label{fig9}
\end{figure*}

Figure \ref{fig8} shows the eigenspectra of the Laplacian matrices for
both the pseudo-fractal (Fig.~\ref{fig8}a) and the Apollonian
(Fig.~\ref{fig8}b) networks, as a function of heterogeneity.
As one sees for $a=2$ (solid lines) synchronizability is observed
only above $\alpha\gtrsim 1.5$, and in particular there is no
synchronizability for the homogeneous coupling regime ($\alpha=0$).
Figure \ref{fig9}a shows the distribution of the average standard
deviation over a sample of $500$ initial configurations, from which
one clearly sees that there are no coherent solutions.
Here the standard mean deviation is characterized by some large value
which is almost constant beyond the weak coupling regime
($\varepsilon\gtrsim 0.2$). 
In the weak coupling regime ($\varepsilon\lesssim 0.2$) the standard
mean square deviation is even larger, since the coupling is not strong
enough to compensate the highly chaotic local dynamics ($a=2$).
%One possible physical explanation for this absence of synchronizability is
%that long range random connections are crucial to improve the ability
%for synchronization and, due to the deterministic construction of the 
%network, there are no long range connections as in the Barab\'asi-Albert 
%scale-free network.
%%%%%%%%%%%%%%%%%%%%%%%%%%%%%%%%%%%%%%%%%%%%%%%%%%%%%%%%%%%%%%%%%%%%%%%
%%%%%%%%%%%%%%%%%%%%%%%%%%%%%%%%%%%%%%%%%%%%%%%%%%%%%%%%%%%%%%%%%%%%%%%
%%%%%%%%%%%%%%%%%%%%%%%%%%%%%%%%%%%%%%%%%%%%%%%%%%%%%%%%%%%%%%%%%%%%%%%
%%%%%%%%%%%%%%%%%%%%%%%%%%%%%%%%%%%%%%%%%%%%%%%%%%%%%%%%%%%%%%%%%%%%%%%
%%%%%%%%%%%%%%%%%%%%%%%%%%%%%%%%%%%%%%%%%%%%%%%%%%%%%%%%%%%%%%%%%%%%%%%
\begin{figure*}
\begin{center}
\includegraphics*[width=7.5cm]{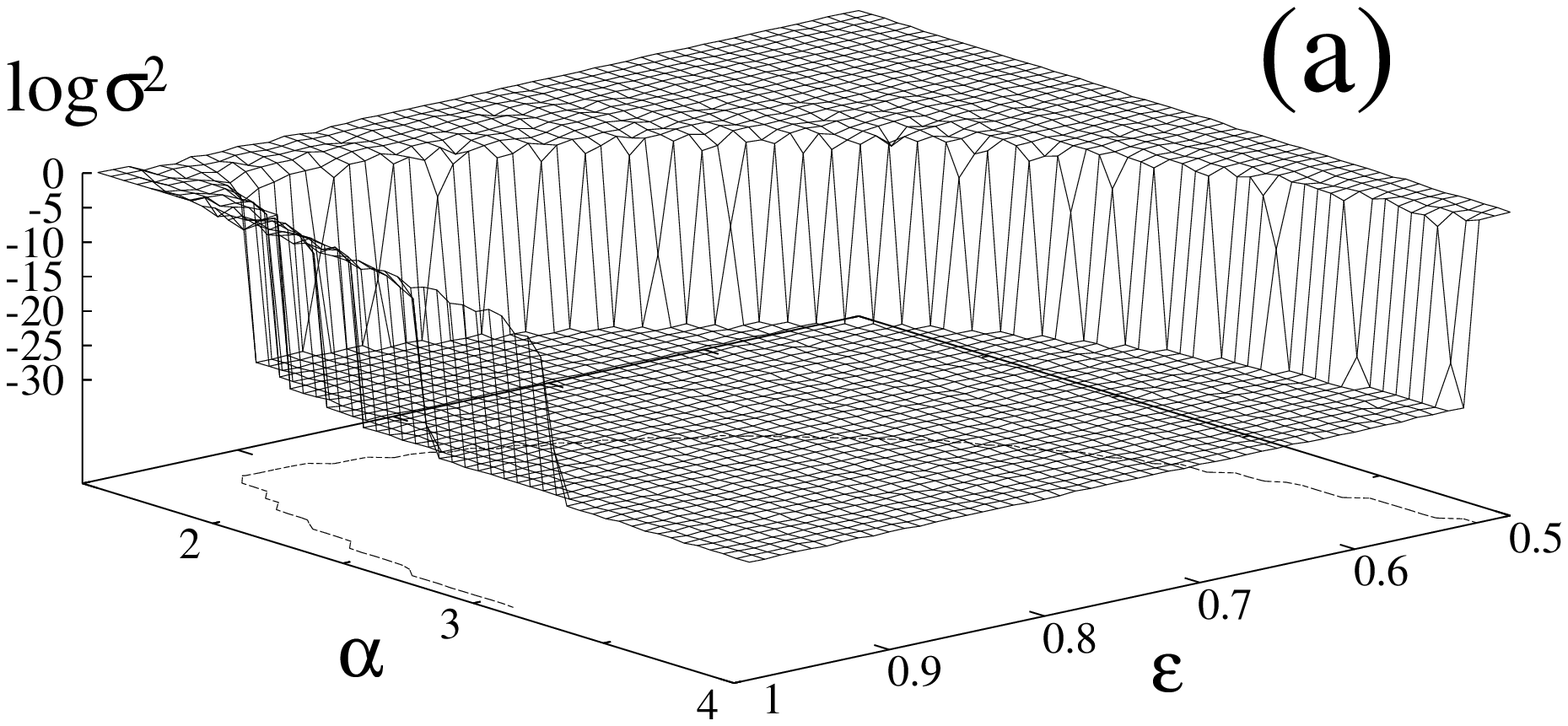}%
\includegraphics*[width=7.5cm]{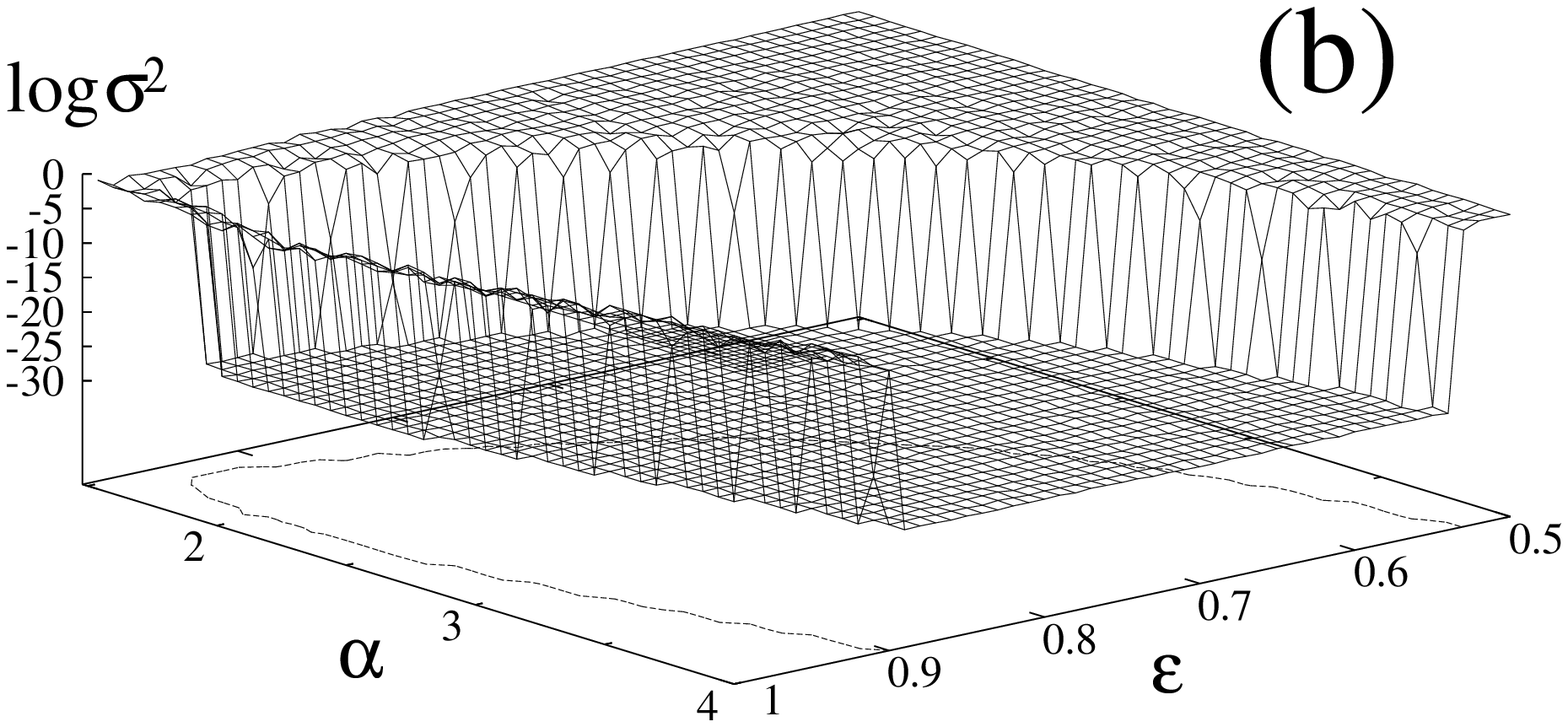}
\includegraphics*[width=12.5cm]{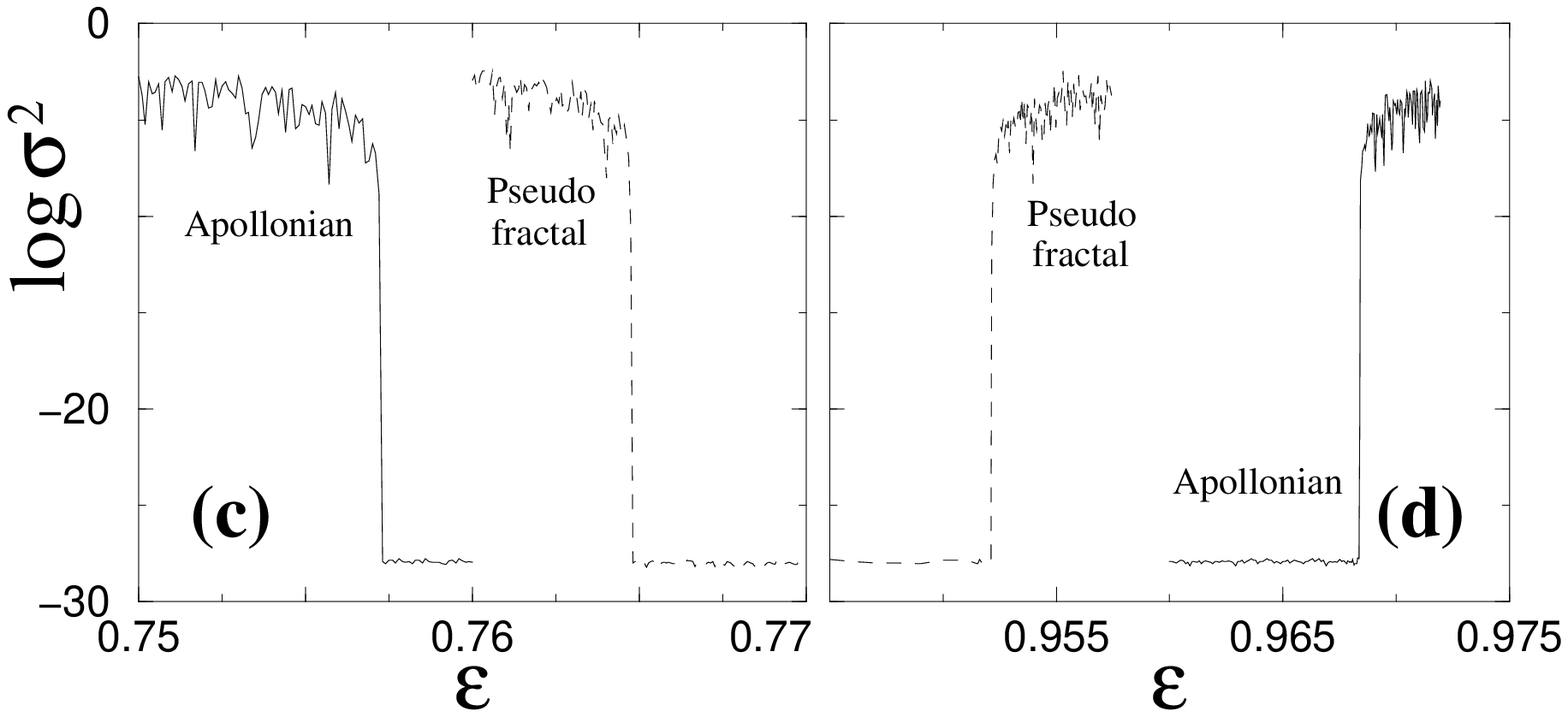}
\end{center}
\caption{\protect 
  Inducing transition to coherence by varying the heterogeneity
  $\alpha$ (see Eq.~(\ref{model})) in scale-free networks.
  {\bf (a)} pseudo-fractal network and
  {\bf (b)} Apollonian network. 
  For strong heterogeneity coherence appears beyond
  a relatively high coupling strength, and
  disappears again for very large couplings (see text).
  For each network, we use $\ell = 6$ generations of nodes and fix
  $a=2$.
  {\bf (c)} and {\bf (d)} show high-resolution plots of $\sigma^2$ as
  a function of $\varepsilon$ for $\alpha=2$, emphasizing the
  first-order phase transition to coherence.} 
\label{fig10}
\end{figure*}
%%%%%%%%%%%%%%%%%%%%%%%%%%%%%%%%%%%%%%%%%%%%%%%%%%%%%%%%%%%%%%%%%%%%%%%
%%%%%%%%%%%%%%%%%%%%%%%%%%%%%%%%%%%%%%%%%%%%%%%%%%%%%%%%%%%%%%%%%%%%%%%
\begin{figure*}
\begin{center}
\includegraphics*[width=15.0cm]{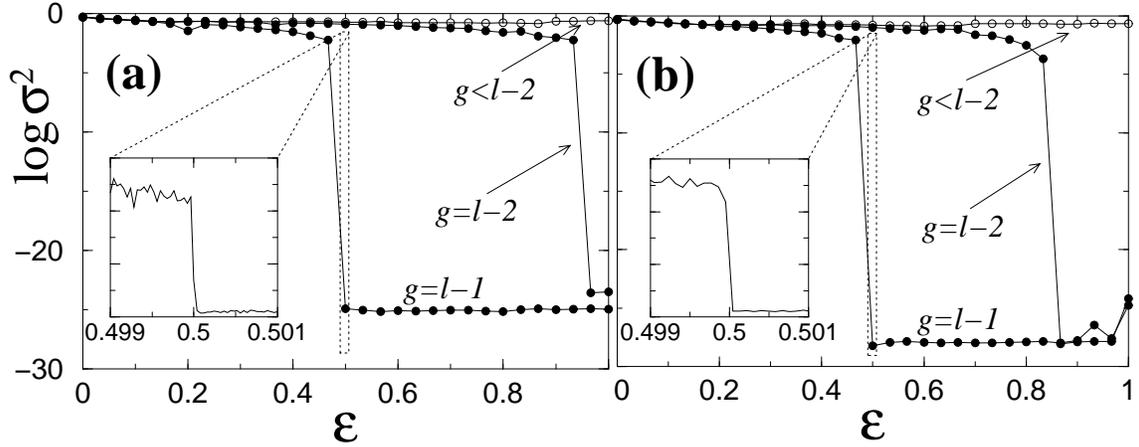}
\end{center}
\caption{\protect 
  Transitions to coherence in deterministic scale-free
  networks, when synchronizing the first $g$ generations of nodes out
  of $\ell$ generations (see text).
  {\bf (a)} pseudo-fractal network and
  {\bf (b)} Apollonian network. 
  The collective dynamical behavior is quite insensitive to hubs (see text).
  Insets show that transitions to coherence are of first-order.
  For each network, we use $\ell = 9$ generations of nodes and $a=2$ fixed.} 
\label{fig11}
\end{figure*}
%%%%%%%%%%%%%%%%%%%%%%%%%%%%%%%%%%%%%%%%%%%%%%%%%%%%%%%%%%%%%%%%%%%%%%%

From Fig.~\ref{fig8} one also sees that, for the pseudo-fractal and
$\alpha > 1.5$, the upper threshold $\varepsilon_U$ increases
monotonically with the heterogeneity, while for the Apollonian network
the upper threshold decreases.
This particular difference between both networks should be due to
their geometrical differences, in particular the fact that 
Apollonian networks are embedded in Euclidean space could explain in
some way that stronger dominance in the coupling to the
most connected nodes {\it destroys} coherence. 

Choosing other values of $a$ for which local dynamics is chaotic, one
finds the same values of $\varepsilon_L$ and $\varepsilon_U$ as
functions of $\alpha$ only shifted: $\varepsilon_L$ gets smaller,
while $\varepsilon_U$ increases.
Figure \ref{fig8} illustrates this for the particular case of $a=1.5$.
Decreasing even further the nonlinearity below the accumulation point
$a=1.411\dots$ synchronizability is attained for any positive value of
$\alpha$, whenever the coupling strength is sufficiently strong. 

Moreover there is a complicated dependence of the average standard
deviation on the coupling strength and nonlinearity.
As shown in Fig.~\ref{fig9}b for deterministic scale-free networks one
finds two main regions in the $(a,\varepsilon)$ plane:
(I) a region where the standard mean square deviation is large and
varies smoothly with the parameters and 
(II) a region where the mean square deviation is smaller but has
larger fluctuations.
This second region, observed for $a\lesssim 1.7$, is somehow
surprising, since irregular variations of the standard mean
square deviation occur for low nonlinearity and high coupling
strengths, precisely where one would expect the most regular behavior
of the node dynamics. 
%As for the exponent in Eq.~(\ref{powerlaw}), it seems that here one
%has also a separation of nontrivial behavior at $a\sim 1.7$. 

As for the heterogeneous coupling regime ($\alpha\neq 0$),
Fig.~\ref{fig10} illustrates the transition to coherence by
varying the heterogeneity $\alpha$ for the pseudo-fractal
(Fig.~\ref{fig10}a) and the Apollonian network (Fig.~\ref{fig10}b).
For both networks, one sees that coherence sets in for $\alpha\gtrsim
1.5$, and the contour of the histograms marking the transition to
coherence fits well the regions in Fig.~\ref{fig8} labeled as 'sync'.
Moreover, from Figs.~\ref{fig10}c and \ref{fig10}d one observes that
all these transitions to coherence are of first-order.

Finally, we study the role of hubs in deterministic scale-free
networks, as we did in the previous Section for random networks.
To this end, we impose synchronization among $g=1,\dots,\ell$
generations, with $\ell$ being the total number of generations,
and observed in what conditions coherent states are observed.
In the pseudo-fractal network the first generation has
$N_1-N_0=3$ nodes, the second one has 
$N_2-N_1=9$ nodes, and the $n$th generation has
$N_{n}-N_{n-1}=3^{n}$ nodes.
In the Apollonian network the number of nodes appearing at each
generation is precisely the same.

Figure \ref{fig11} shows the standard mean square deviation as a
function of coupling strength for pseudo-fractal (Fig.~\ref{fig11}a)
and Apollonian networks (Fig.~\ref{fig11}b).
In each case we choose the fully chaotic map ($a=2$) and impose 
synchronization among the nodes of the first $g$ generations
by setting them to be their mean amplitude at each time-step.
In both cases, one sees that the standard mean square deviation
remains large when synchronization is imposed to all $g<\ell-2$
generations. 
Coherent solutions are only observed for $g=\ell-2$ and $g=\ell-1$,
beyond a coupling threshold which is smaller for the latter
case. 
Surprisingly, for $g=\ell -1$ the transition to coherence occurs
for the same coupling strength in both networks.
This may be due to the fact that, the fraction
$N_g/N_{\ell}=3^{g-\ell}=1/3$ of nodes on which one imposes
synchronization is the same for both networks and is high enough to
suppress the influence of local connectivities.

For $g=\ell-2$ the pseudo-fractal network shows coherence
only above very high coupling strengths, near $\varepsilon\sim 1$,
while for Apollonian networks the threshold is much lower.
This difference in the coupling strength threshold is due to the
fact that here the fraction of nodes $N_g/N_{\ell}=1/9$ to which
synchronization is imposed is small enough to not suppress the
influence of local connectivities. 
Therefore, since the hubs in the pseudo-fractal network are less
connected than the hubs in Apollonian networks, one needs higher
coupling strength to observe coherence.
For any higher value $\ell$ of generations the same results are
obtained, since one has for the quotient of the number  of nodes
between two successive generations $N_n/N_{n-1}\to 3$ as $n$
increases. 

As a general remark, one observes from Fig.~\ref{fig11} that one needs
to synchronize a rather high fraction of nodes ($\gtrsim 1/9$) 
to induce coherence. 
Therefore, it seems that, dynamical collective behavior on scale-free
networks is quite insensitive to hubs. 
As shown in the insets of Figs.~\ref{fig11}a and
\ref{fig11}b, the transition to coherence is also of first-order.

%%%%%%%%%%%%%%%%%%%%%%%%%
\section{Discussion and conclusions}
\label{sec:discussion}

In this paper we studied fully synchronized solutions for three
scale-free network topologies.
The main conclusion is the following:
in random scale-free networks synchronization of chaotic maps not only
depends on the coupling strength but is mainly controlled by the
outgoing connectivity $k$, which is a measure of cohesion in the networks.
Because of that, one finds coherent solutions in random scale-free
networks of fully chaotic logistic maps ($a=2$) with outgoing
connectivity $k=8$ and homogeneous coupling, but not in deterministic
scale-free networks, since they have rather small effective outgoing 
connectivity, namely $k=2$ for the pseudo-fractal network and $k=3$
for the Apollonian network.
Therefore, although the exponent $\gamma$ of connection distributions
in scale-free networks does not depend on the outgoing connectivity 
\cite{albert02}, we have shown that, in general, synchronization of
chaotic maps in such coupling topologies is quite sensitive to it. 

Our results were obtained both numerically, from histogram of
significantly large samples of initial configurations with a criterion
for full synchronization based on the mean standard deviation of
amplitudes, Eq.~(\ref{sigma}), and analytically from the eigenvalue
spectra of the diagonalized variational equations computed at the
coherent states, Eq.~(\ref{linestab}). 

In particular, for random scale-free networks, the threshold values of
the coupling strength obey a power-law, Eq.~(\ref{powerlaw}), as
function of the outgoing connectivity. 
The exponent of this power-law depends on the nonlinearity $a$ of the
chaotic map, being almost constant below $a_c\sim 1.7$ and decreasing
linearly above it. Interestingly this value of $a_c$ is in the
vicinity of the bifurcation of the quadratic map where the period-$3$
window appears, and coincides with the appearance of other
nontrivial behaviors in coupled map lattices with regular topologies,
namely in the velocity distribution of traveling wave solutions
\cite{lind04b}. 

For deterministic scale-free networks with homogeneous coupling, the
same value $a_c$ indicates the threshold above which no coherent
solutions are observed, independently of the coupling strength.
Above $a_c$, coherence is observed only for heterogeneous
coupling, namely for $\alpha\gtrsim 1.5$.
However, for this range of values, we have also shown that coherence is  
also absent either for very small or for very large coupling
strengths, due to spatial instabilities.
Another particularly interesting result that still needs to be
explained is that, for Apollonian networks, the coupling threshold
beyond which coherence disappears gets smaller when the
heterogeneity is further increased. 
This point is not observed for the pseudo-fractal network and may be
due to the geometrical differences between both deterministic
networks.

As a general property, we have shown that all transitions to
coherence are of first-order, indicating a similarity with other
complex networks \cite{strogatz01}. 
Furthermore, all results are robust not only against changes of the
initial configurations of node amplitude but also, in random
scale-free networks, against changes of the connection network. 
We also presented results indicating that in scale-free networks hubs
play apparently no fundamental role in the dynamical collective
behavior. 

%%%%%%%%%%
\section*{Acknowledgments}

The authors thank A.O.~Sousa and C.~Zhou for useful discussions. 
P.G.L.~thanks Funda\c{c}\~ao para a Ci\^encia e a Tecnologia, Portugal, for
financial support. 
J.A.C.G.~thanks Conselho Nacional de Desenvolvimento Cient\'{\i}fico e
Tecnol\'ogico, Brazil  and Sonderforschungsbereich 404 of DFG for
financial support.  

%%%%%%%%%%%%%%%%%%%%%%%%%%%%%%%%%%%%%%%%%%%%%%%%%%%%%%%%%%%%%%%%%%%%%%%%%%%
%%%%%%%%%%%%%%REFERENCIAS %%%%%%%%%%%%%%%%%%%%%%%%%%%%%%%%%%%%%%%%%%%%%%%%%
%%%%%%%%%%%%%%%%%%%%%%%%%%%%%%%%%%%%%%%%%%%%%%%%%%%%%%%%%%%%%%%%%%%%%%%%%%%

%%%%%%%%%%%%%%%%%%%%%%%%%%%%%%%%%%%%%%%%%%%%%%%%%%%%%%%%%%%%%%%%%%%%%%%%%%%
\end{document}